\documentclass[journal]{IEEEtran} 
\usepackage{amsmath,amssymb}
\usepackage{amsfonts}
\usepackage{graphicx}
\usepackage{cite}
\usepackage{pgf}
\usepackage{tikz}
\usepackage{rotating}
\usepackage{setspace}
\usepackage{color}
\usetikzlibrary{arrows,automata}
\usetikzlibrary{calc}

\usepackage{fancyhdr}
\usepackage{textcomp}
\newcommand{\blue}[1]{#1}
\newcommand{\red}[1]{#1}
\newcommand{\grn}[1]{#1}
\newcommand{\bII}[1]{#1}
\newcommand{\grII}[1]{#1}
\newcommand{\med}[1]{#1}
\newcommand{\figtop}{\vspace{-0.4cm}}
\newcommand{\figtopsmall}{\vspace{-0cm}}
\newcommand{\figbottom}{\vspace{-0.5cm}}

\newtheorem{lemma}{Lemma}

\newcommand{\beq}{\begin{equation}}
\newcommand{\eeq}{\end{equation}}
\newcommand{\beqa}{\begin{eqnarray}}
\newcommand{\eeqa}{\end{eqnarray}}
\newcommand{\pkt}{\textbf{p}}
\newcommand{\spn}{\mathrm{span}}
\newcommand{\mubar}{\overline{\mu}}
\newcommand{\lambar}{\overline{\lambda}}
\newcommand{\lambde}{\lambda_{\mathrm{est}}}

\newcommand{\imgsize}{0.95\columnwidth}

\abovecaptionskip
\belowcaptionskip

\begin{document}

\title{Dynamic Rate Adaptation for Improved Throughput and Delay
 in Wireless Network Coded Broadcast}
\author{\IEEEauthorblockN{Amy Fu, Parastoo Sadeghi and Muriel M{\'e}dard }
\thanks{A. Fu and P. Sadeghi are with the Research School of Engineering, College of
Engineering and Computer Science, The Australian National
University, Canberra, Australia. Emails:
\{amy.fu;parastoo.sadeghi\}@anu.edu.au.

M. M{\'e}dard is with the Department of Electrical Engineering and
Computer Science, Research Laboratory of Electronics, Massachusetts
Institute of Technology, Cambridge, MA 02139 USA. Email:
medard@mit.edu. }} \maketitle
\begin{abstract}
In this paper we provide theoretical and simulation-based study of
the \blue{delivery} delay performance \blue{of} a number of existing throughput optimal
\blue{coding} schemes and use the results to design a new dynamic rate
adaptation scheme that achieves \blue{improved overall throughput-delay performance.}

Under \blue{a baseline rate control scheme}, the receivers' delay performance is examined. Based on their Markov states,
the knowledge difference between the sender and receiver, three
distinct methods for \blue{packet delivery} are identified: zero state, leader
state and coefficient-based \blue{delivery}. We provide analyses of each of
these and show that, in \blue{many} cases, zero state \blue{delivery} alone presents
a tractable approximation of the expected packet delivery behaviour.
Interestingly, while coefficient-based \blue{delivery} has so far been
treated as a secondary effect in the literature, we find that the
choice of coefficients is extremely important in determining the delay, and a well chosen encoding scheme can, in fact, contribute a significant improvement to the delivery delay.

Based on our delivery delay model, we develop a \blue{dynamic rate adaptation scheme which uses performance prediction models to determine the sender transmission rate}. \grII{Surprisingly, taking this approach leads us to the simple conclusion that the sender should regulate its addition rate based on the total number of undelivered packets stored at the receivers. We show that despite its simplicity,} our proposed dynamic rate adaptation scheme results in noticeably improved throughput-delay performance \blue{over existing schemes in the literature}.

\end{abstract}

\section{Introduction}\label{sec:intro}
In recent times there have been many advances in the capabilities of wireless
communication \blue{systems \cite{fragouli08}. A number of applications can now take advantage of
these new capabilities, requiring high data rate and low delay performance. In
this paper, we consider applications in which the same ordered set of packets is
required by all receivers, with low delay. One application is video
broadcasting where the receivers not only wish to watch live video, but also
want to keep a high quality copy for later use. This might include the
broadcast of a lecture or conference recording, or perhaps simultaneous
download and viewing of a purchased movie. Other potential applications include
the broadcast of common information in multiplayer gaming, where users' actions
must be logged in order, as well as certain scientific or mission-critical
applications with low delay requirements. This differs from work such as \cite{chou06,seferoglu09,chakareski06}, in that the applications considered in this paper do not tolerate packet losses.}

In \blue{the context of these applications} there are two key measures of
performance. \blue{One measure is \emph{throughput}, defined as the
average rate at which packets are delivered across receivers. This measures the efficiency with which the receivers' channel bandwidth is utilised. }
Since packets can only be \blue{used} in order, we can only consider a packet useful once it has been \emph{delivered}, that is if it and all preceding packets have also been correctly received. Low delay is also desirable, to avoid \blue{latency at the application}. Therefore, it is \blue{equally} important to minimise the \emph{delivery delay}, the \blue{average} time between when a \blue{packet is first available for transmission} to the time it is delivered to \blue{the application layer}.

Meeting these requirements in a wireless setting is not an easy task
\cite{fragouli07}. Receivers' independent channel conditions mean that they will experience very different erasure patterns, which in turn leads to a variety of \blue{packet} demands on the sender.

\subsection{Network coding} \label{sec:introNC}
\emph{Linear network coding} \cite{ahlswede1,koetter1,fragouli07} \blue{is used
as} an effective way to accommodate multiple receivers' \blue{packet demands}
while still efficiently using the transmission bandwidth. Under
\blue{linear} network coding the sender divides the \blue{information} into
equal sized packets, and combines a number of packets into each transmission
using Galois field arithmetic \cite{koetter1}. This combination is
\blue{transmitted} to the receivers along with the \red{\emph{coefficients}} used \blue{to
combine the packets}. In order to recover the original packets, receivers must
collect enough \blue{coded} packets to decode them using Gaussian
elimination \cite{koetter1}.


Although network coding \blue{is known to enhance the throughput in many
networks}, the time spent waiting to receive the necessary packet combinations
for decoding \blue{can result} in an additional \emph{decoding delay}.
\blue{There are two problems associated with a large decoding delay. Firstly,
the decoding delay} lower bounds the achievable delivery delay since packets
can only be delivered after being decoded. \blue{Secondly, undecoded
packets can greatly increase the computational complexity of
operations for both the sender and receiver. \red{Gaussian
elimination, required for receivers to decode, is known to} scale as the cube of the number of
packets in the set. In \emph{full feedback} systems the sender
performs similar operations to determine what information is
missing at the receivers. Large decoding delays mean that,
on average, \med{many} undecoded packets will be stored at the
receivers, resulting in more computationally expensive packet
transmissions.}

\blue{Network coding introduces \red{a well known tradeoff}\cite{nguyen1,li3,barros1,zeng12,eryilmaz08} between throughput and delivery delay. \red{Generally the more stringent the delay requirements, the more throughput must be sacrificed to achieve them. Many transmission schemes have been devised} that aim at striking a balance between high
throughput and low delay in network coded systems. We will present an overview of existing approaches in the literature, and highlight the open question\red{s} that will be addressed in this paper. For brevity, we will focus on
broadcast applications in wireless packet erasure channels, as they are directly
related to our work.}

\subsection{Existing methods for delay control}
\red{To ensure that packets can be delivered in a timely fashion, it is necessary to introduce some controlled redundancy into the sender's transmissions. This allows receivers who have experienced channel erasures to recover and deliver their missing packets. The \emph{transmission schemes} used to achieve this can generally be divided into two components: a \emph{rate control scheme} and a \emph{coding scheme}. More detail will be provided in Section \ref{sec:model}. Essentially, the rate control scheme determines the \emph{transmission rate}, the number of new packets that can be included in the sender's transmissions at each time, while the coding scheme is responsible for determining the coefficients. Each of these components can have an impact on the throughput and delay.}

\subsubsection{Rate control}\label{sec:introratectrl}
\red{There are a number of ways to use rate control to reduce the delivery delay.}

\red{Under \emph{block based} transmission schemes,} \blue{incoming packets are divided into blocks or generations \cite{chou2,swapna10,vehkapera05,park1,eryilmaz08,lucani2,lun06,gkantsidis2,sadeghi1, sorour1,sadeghi3,li11,sorour10,yang12}. \red{The rate control scheme only allows the packets of one block to be transmitted at a time, ensuring that} a block's worth of \emph{innovative information} has been received by every receiver, \red{before moving} on to the next block. The primary advantage of this rate control scheme is that \red{since} packet delivery is done on a block-by-block basis, shorter block lengths mean smaller delivery delays. However, this comes at the cost of lower (even vanishing) throughput \cite{swapna10}. \red{Another advantage of} this rate control scheme is that it requires \red{only} minimal
feedback from the receivers about block completion \cite{lucani2,zeng12,fragouli2}.}

\red{Other transmission schemes such as \cite{sundararajan2,sundararajan3,barros1} are non-block based. In \cite{barros1} a rate control scheme is implicitly implemented where new packets may be transmitted only if the delay performance, determined from receiver feedback, is sufficiently good. In contrast, \cite{sundararajan2,sundararajan3} make little use of feedback in determining the transmission rate. Instead they use a fixed transmission rate, and rely on natural fluctuations in the \emph{transmission queue} size to ensure the delivery of packets. While there has been much work studying the delay performance of block-based transmission schemes \cite{swapna10,lucani2,park1,chou2,eryilmaz08,vehkapera05}, so far only asymptotic limits \cite{sundararajan1} on the delay performance of \cite{sundararajan2,sundararajan3} have been found.}

\subsubsection{Coding} \label{sec:redcod}
\red{The coding scheme may be used to further improve the delay performance.}

\red{Under some coding schemes, the sender transmits network coded packets which may be noninnovative to selected receivers. A good example of this is} \blue{\emph{instantaneously decodable} network coding \cite{sadeghi1,sorour1,sadeghi3,li11,sorour10}. In this \red{block-based transmission scheme, feedback about packets stored at the receivers is used to construct transmissions that allow immediate decoding at a subset of (or if possible all) receivers. However instantaneous decodability comes at the cost of reduced throughput, since} not every receiver may receive \red{innovative information} in every transmission.}

\red{By contrast \emph{throughput optimal} coding schemes do not attempt to introduce more redundancy, but instead aim to maximise the number of receivers that can obtain innovative information from each transmission. Random linear network coding (RLNC), where coefficients are chosen at random, is the most common. In \cite{ho1} this was shown to achieve the capacity of a multicast network with high probability as the field size becomes large. The simplicity of implementation has led to a great deal of work including \cite{lun06,swapna10,lucani2,gkantsidis2,park1,chou2,eryilmaz08,vehkapera05}.}

\red{Feedback-based throughput optimal coding schemes have also been proposed to reduce the transmission queue size \cite{sundararajan3}, and minimise the delivery delay \cite{sundararajan2}, however no attempt has been made to study the extent to which these schemes work.}

\blue{To the best of our knowledge, 1) there has been no work on characterizing the
non-asymptotic \red{delivery} behaviour of the rate control scheme used in \cite{sundararajan3,sundararajan2}, \red{2) the delay performance of the coding schemes presented in \cite{sundararajan2,sundararajan3} \grn{has} not yet been analysed, and 3)} there has been no} \red{systematic attempt to implement a rate control scheme that adaptively considers both the throughput and delay performance in determining the transmission rate.}

\subsection{Contributions and distinctions with related work}

In this paper we take a first step in realising a dynamic tradeoff between
throughput and \blue{delivery} delay in a wireless \blue{network coded} broadcast system. By first
understanding the mechanism by which packets are delivered in
\blue{transmission} schemes such as \cite{sundararajan3,sundararajan2}, we gain
insight into the nature of the \red{\emph{throughput-delay tradeoff}, the set of throughput values and delivery delays simultaneously achievable by a system}. This in itself is a
difficult problem, owing to the complex interactions between the sender and
receivers. To manage this, we categorise the methods of \blue{packet} delivery into three
categories: zero state, leader state and coefficient-based \blue{delivery}. These
distinctions are made on the basis of receivers' Markov states: defined as the
difference between \blue{the number of packets known by the} sender and
receiver at each time step. By decoupling the contributions of each method of
delivery, we can present an approximation that removes the effect of cross
receiver interactions. In return for some loss of accuracy, we are able to
transform a mathematically intractable problem into one that gives
easily calculable results.

Based on our understanding of the mechanics of broadcast packet delivery, we
propose a new transmission scheme which uses feedback information to
\blue{predict the receivers' short term throughput and delivery performance. This is then used to determine when to \red{include} new packets \red{into the sender's transmissions}. In effect, the sender dynamically tailors the transmission rate for noticably
improved throughput-delay performance compared with
\cite{sundararajan2,sundararajan3,barros1}}. A related idea is considered in
\cite{yang12}, where the \red{block size} is chosen to maximise the number of packets delivered to all receivers by a hard deadline. However, \red{as commented in Section \ref{sec:introratectrl}} block coding is generally not conducive to good throughput.

\section{System Model}\label{sec:model}

\blue{\grn{A single sender aims to transmit a backlogged set of} data packets $\pkt_1,\pkt_2,\cdots$ in the correct order \grn{to a} set of $R$ receivers}. \blue{Time is slotted, denoted by $t =
1, \cdots,$ and the} sender can broadcast at the rate of one
\blue{original or network coded} packet per time slot. The receivers are
connected to the sender via independent erasure channels \blue{with \emph{channel rate} $\mu$, so that they} successfully receive transmissions with probability $\mu$ at each time
slot.\footnote{In general receivers may have different channel \blue{rates}, but for clarity of explanation we only consider the homogeneous
case.}

Receivers store received \blue{packets} in a buffer and send an
acknowledgement after each successful \blue{packet} reception \blue{or a
negative acknowledgement if the packet is discarded owing to an erasure}, which
we assume the sender detects without error.\footnote{Although this \blue{can be difficult to achieve in practice}, it greatly simplifies analysis. We will make some
comments on the effect of imperfect feedback later in this work.} The sender uses this information to record which packets receivers have stored in their buffers. \blue{Based on this information, a \emph{transmission scheme} can be devised to determine the packet combinations the sender will transmit. The components of the transmission schemes we will study will be outlined in the remainder of this section. We now define the \emph{delivery delay} and \emph{throughput}, which will be used to compare the performance of the transmission schemes studied in this paper.}

\subsubsection{Packet delivery} At time slot $T$, a packet $\pkt_n$ is said to be \emph{delivered} to a receiver if that receiver has already decoded all packets $\pkt_1,..., \pkt_{n-1}$ and \blue{first} decodes $\pkt_n$ at \grn{time} $t = T$. Otherwise, $\pkt_n$ is said to be \emph{undelivered} to that receiver. 

\subsubsection{Delivery delay} \blue{The delivery delay of a transmission scheme is measured as the average} number of time slots \blue{between any packet $\pkt$ becoming available for transmission, to the time it is delivered to each of receivers.}

\subsubsection{Throughput} \blue{The throughput of our system is measured as the average number of packets delivered per time slot, across receivers.}
%

\subsection{Transmission scheme}

\begin{figure*}
\figtop
\begin{center}
\begin{tikzpicture}[auto]
 \tikzstyle{block} = [rectangle, draw, text width=5.5em, text centered,minimum height=3em]
 \tikzstyle{block2} = [rectangle, draw, text width=5em, text centered,minimum height=1.8em,node distance=0.75cm] 
 \tikzstyle{box}=[rectangle,draw,minimum height=4em,text width=15em, text top]
 \tikzstyle{line} = [draw, -latex']
 \node[block](arriv){Packet backlog};
 \node[block](ratectrl)[right of=arriv,node distance =3cm]{Rate control scheme};
 \node[block](txq)[right of=ratectrl,node distance=3cm]{Transmission queue};
 \node[block](cod)[right of=txq,node distance =3cm]{Coding scheme};
 \node[block2](rx2)[right of=cod,node distance =5.5cm]{$\cdots$};
 \node[block2](rx3)[below of=rx2]{Receiver $R$};
 \node[block2](rx1)[above of=rx2]{Receiver 1};
 \path [line] (arriv) -- (ratectrl); 
 \path [line] (ratectrl) -- (txq);
 \path [line] (txq) -- (cod);
 \path [line, dashed](cod) edge node[rectangle, text width=8em, minimum height=6.5em, text centered] {Wireless packet erasure channels} (rx2);
 \path [line,dashed] (cod)--(rx1);
 \path [line,dashed] (cod)--(rx3);
  \draw[black,thick,dotted] ($(ratectrl.north west)+(-0.2,0.2)$)  rectangle ($(cod.south east)+(0.2,-0.2)$);
	\node (txscheme) [above of=txq] {Transmission scheme};

\end{tikzpicture}
 \caption{A block diagram of the components of a transmission scheme.} \label{fig:txscheme} \
 \figbottom
\end{center}
\end{figure*}
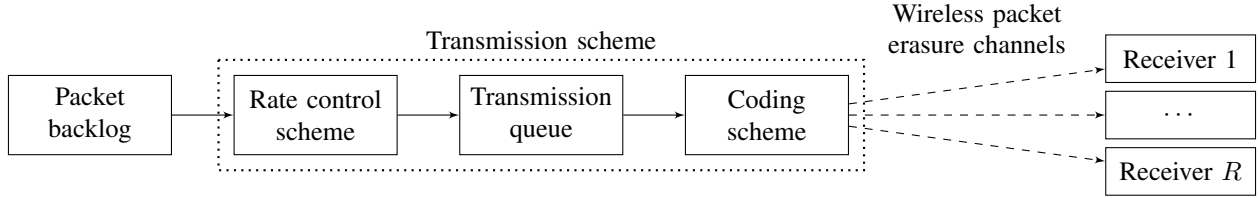

\blue{Here we outline the model for the transmission schemes we will be studying, as shown in Fig. \ref{fig:txscheme}. The transmission scheme employed by the sender can be divided into three components: a \emph{rate control block}, which passes new packets into a \emph{transmission queue}, from which a \emph{coding block} determines $c(t)$, the network coded transmission to be sent at time $t$. We briefly outline the function of each block here.}
\subsubsection{Rate control block}
\blue{The rate control block employs a \emph{rate control scheme} to decide when to introduce new packets from the application into the transmission queue. In our paper, we assume the application has an infinite backlog of packets
available to be transmitted by the sender. Since the sender transmits one packet per time slot, we limit the rate control scheme to pass at most one new packet per time slot to the transmission queue. Therefore at each time $t$, the rate control block can decide whether to \emph{add}, and place a new packet in the transmission queue, or \emph{wait} and do nothing. If the rate control block adds, then we set the \emph{add decision} $a(t)=1$; if it waits, $a(t)=0$.}
\subsubsection{Transmission queue}
\blue{The transmission queue stores all packets passed by the rate control scheme. Only packets in the transmission queue may be transmitted by the sender.} Once all receivers have decoded a packet $\pkt$, it is removed from the \blue{transmission queue.\footnote{In \cite{sundararajan3}, packets may be removed from the transmission queue before they are decoded by all receivers. However we ignore this option, as transmission schemes other than \cite{sundararajan3} are also considered \med{and we do not explicitly attempt to manage the queue size}.} At any time $t$, the total number of packets that have been passed into the transmission queue is}
\beq A(t)=\sum_{i=1}^t a(i). \label{eq:totarriv}\eeq
\blue{The remainder of this paper will focus on delivering the packets in the transmission queue to the receivers. Therefore, with a slight abuse of notation, the packets in the transmission queue will be referred to, from oldest to newest, as $\pkt_1,...,\pkt_{A'(t)}$, where $A'(t)$ is the total number of packets in the transmission queue at time $t$.}

\subsubsection{Coding block} \label{sec:codblock}
\blue{The coding block \grn{employs a \emph{coding scheme} to} determine which of the packets in the transmission queue to code into the outgoing transmission \blue{$c(t)$} at each time slot. Since the coding \blue{block may only choose packets from the transmission queue}, \grn{transmissions are} of the form
 \beq c(t)=\sum_{i=1}^{A'(t)} \alpha_i(t) \pkt_i, \label{eq:cod}\eeq
where the coefficients $\alpha_i(t)$ are chosen at each time slot from the
field $\mathbb{F}_M$ of an appropriate size.\footnote{The coding schemes of \cite{sundararajan3,sundararajan2} prove that it is always possible to find an \emph{innovative} combination for all receivers if $M\geq R$, the number of receivers.
}} This combination is transmitted
along with the corresponding \blue{\emph{transmission vector}} $v_s(t)$. If each uncoded
packet $\pkt_i$ corresponds to the standard basis vector $e_i$ whose $i$-{th}
entry is $1$, then
\begin{eqnarray} v_s(t)&=& \sum_{i=1}^{\blue{A'(t)}}\alpha_i(t)e_i \nonumber \\
&=&\left[\alpha_1(t),\alpha_2(t),\grn{\cdots} \right] \label{eq:txvector}
\end{eqnarray}
so that the $i$-th entry \blue{of the transmission vector} $\alpha_i(t)$ corresponds to the coefficient of $\pkt_i$. Receivers use \red{the information in the transmission vector} to recover the original packets by performing Gaussian elimination on the
packets in their buffers.

\subsection{Rate control schemes}\label{sec:arrival}

\blue{In this paper we consider three rate control schemes. \grn{Two of these, the \emph{delay threshold} and \emph{dynamic} rate control schemes, are both} \emph{rate adaptation schemes}, which use feedback from the receivers to adjust their transmission rates. As a means of comparison we will also study a \emph{baseline} rate control scheme, which does not utilise feedback from the receivers \grn{to determine the transmission rate}.}
\subsubsection{\bf \blue{Baseline rate control scheme}}
\blue{Under this rate control scheme, the add decision $a(t)$ is} determined by a Bernoulli process with \blue{\emph{addition rate} $\lambda$, so that the sender will add with probability} $\Pr(a(t)=1)=\lambda$ independently at each
time slot $t$. \blue{This is equivalent to the model used in \cite{sundararajan3,sundararajan2}.\footnote{To be precise, in \cite{sundararajan3,sundararajan2} packets are assumed to \emph{arrive} at the application by a Bernoulli process. We have transformed this into the equivalent rate control scheme to make it comparable in terms of throughput and delay to the backlogged schemes studied in this paper.}} By assuming the load factor $\rho=\lambda/\mu$ is appreciably
less than 1, we can provide more practical nonasymptotic analysis of throughput-delay performance.

\subsubsection{\bf Delay threshold scheme}
\blue{This delay threshold rate control scheme is taken from \cite{barros1}, and will
be used as a comparison rate control scheme. This scheme operates under two
modes, which we call \emph{start} and \emph{stop}. By default the sender is set to start mode, where it adds whenever one of the receivers has decoded all packets in the \red{transmission} queue. However if any of the packets inside the transmission queue have
been present for more than some threshold $T_D$ number of time slots,
the sender switches to stop mode. In this case the sender waits, and the coding block transmits uncoded copies of the expired packet(s). Once all packets remaining in the transmission queue are less than $T_D$ time slots old, the sender reverts back to start mode.}

\subsubsection{\bf Dynamic rate control scheme}
\blue{In this paper, we will present a rate control scheme which outperforms
both the baseline and delay threshold rate control schemes. In Section \ref{dynamicscheme} we \med{shall} show how} add and wait decisions \blue{can} be determined \blue{using a \red{delivery} model based on our transmission scheme analysis}.

\subsection{Coding schemes} \label{sec:coeff}

\blue{In this section we outline the three throughput optimal coding schemes we will study in this paper.} To highlight the effects of coefficient selection on delay, we will focus on
two existing schemes, the \emph{drop-when-seen} coding scheme of
\cite{sundararajan3,sundararajan1} and the asymptotically optimal delivery
scheme of \cite{sundararajan2}, which we call \grn{coding} schemes A and B respectively. As
a means of comparison, we also consider a random linear network coding (RLNC) scheme. \blue{Throughput optimal coding schemes all have the \emph{innovation
guarantee property}. This means that, at each time slot, the transmitted packet $c(t)$ will be innovative \med{for} all receivers who are still missing packets in the transmission queue.}
 The method for selecting coefficients in each scheme is summarised
below. More details can be found in \cite{sundararajan3,sundararajan2}.

\subsubsection{\bf Coding scheme A}
Coding scheme A relies on the concept of seen packets. A packet $\pkt_i$ is
\emph{seen} by a receiver if it can use the packets in its buffer to create a
combination of the form $\pkt_i+f(\pkt_{>i})$, where $f(\pkt_{>i})$ is some
linear combination of the \blue{packets} $\pkt_{i+1},\pkt_{i+2},...$. If this is not
possible, then $\pkt_i$ is \emph{unseen}.

Coding scheme A ensures that with each successful \grn{packet} reception, a receiver sees \med{its}
next unseen packet. To determine what \blue{coded} packet to transmit next, the sender lists the
oldest unseen packet from each receiver. Moving from oldest to newest unseen
packet, it adds an appropriate multiple of each packet, \med{so that the resulting packet} is
innovative to all the corresponding receivers.

\subsubsection{\bf Coding scheme B}
Under coding scheme B, the sender transmits a minimal combination based on the oldest
undecoded packet of each receiver. The sender lists these oldest undecoded
packets and their corresponding receivers, then beginning with the newest
packet in the list, it adds in older packets only if the receiver(s) that
correspond to them would not otherwise receive an innovative packet.

\subsubsection{\bf RLNC scheme}
The sender transmits a random combination of all $A'(t)$ packets \blue{in the transmission queue}. Coefficients are selected randomly from $\mathbb{F}_M$. For a fair comparison, feedback is used to ensure that the final packet satisfies the
innovation guarantee property. If the current set of coefficients do not have
this property, new random coefficients are chosen until an appropriate
combination is found.

It should be noted that, unlike the RLNC scheme, \blue{coding schemes A and B are more selective about the packets coded into each transmission. Since} \grn{coding} schemes A and B only code
the oldest unseen or undecoded packet of each user, the sender will only code a new packet
$\pkt_i$ if one of the receivers has decoded $\{\pkt_1,...,\pkt_{i-1}\}$.  As a result, \grn{under these coding schemes} the sender
codes packets from an \blue{\emph{effective transmission queue} which is limited to
the \blue{next needed packet} $\pkt_n$ of the receiver with the most \blue{delivered} packets. \blue{The role of the effective transmission queue will be further discussed in Section \ref{sec:effmarkov}.}}

\subsection{Baseline and coding scheme B transmission schemes}\label{sec:txschemes}
\blue{A transmission scheme is determined by the pairing of a rate control scheme with a coding scheme. In practice, any combination is allowed, however, to simplify the presentation of this paper, two groups of transmission schemes will be studied. In Sections \ref{sec:markov} to \ref{coeffbased}, we analyse the \emph{baseline transmission schemes}: transmission schemes which substitute the baseline rate control scheme into the rate control block. The baseline rate control scheme is chosen as it is the only rate control scheme for which a mathematically tractable model is possible. It is paired with throughput optimal coding schemes A, B and RLNC. In Sections \bII{\ref{dynamicscheme} and} \ref{sec:performance}, we study \emph{coding scheme B transmission schemes}, which substitute coding scheme B into the coding block. Coding scheme B is chosen as it has the best delay performance of the three coding schemes. It is paired with each of the baseline, delay threshold and dynamic rate control schemes.}
\section{Markov State}\label{sec:markov}

Our \blue{delivery delay} analysis will be based on the receivers' \emph{Markov
states,}\footnote{\blue{The Markov state is based on the concept of \emph{virtual queue length} in \cite{sundararajan1}.}} a concept we will explain next. This allows us to
categorise \blue{packet delivery} methods and gives us an important tool for \grn{the}
estimation of the \blue{receivers' delivery delays}.

\subsection{\blue{Knowledge spaces and the Markov state}} \label{sec:kspace}
At time $t$, the \emph{transmission list} is defined
as the set of \blue{standard basis vectors} $V_s(t)=\{e_1,e_2,...,e_{A'(t)}\}$
corresponding to the uncoded packets $\pkt_1,\pkt_2,...,\pkt_{A'(t)}$
which \blue{are currently in the transmission queue}. The sender chooses packets for
transmission from \blue{the \emph{transmission knowledge space}}
 \beq K_s(t)=\spn (V_s(t)), \label{eq:kspace} \eeq
which is the set of all linear combinations the sender can compute \blue{using packets from the transmission queue}.
The size of the \blue{transmission} knowledge space is given by
 \beq |K_s(t)|= M^{|V_s(t)|}, \eeq
where $M$ is the field size and the notation $|X|$ represents the
\med{cardinality} of the set $X$. The \emph{reception list} \red{$V_r(t)$} of a receiver $r$ is defined as the set of received \blue{transmission} vectors which, after Gaussian elimination, correspond to vectors from the current transmission knowledge space $K_s(t)$. The \blue{\emph{receiver knowledge space} is similarly defined as the set of all linear combinations that can be calculated from its reception list,} $K_r(t)=\spn (V_r(t))$. These concepts will be used in our analysis of coefficient-based delivery in Section \ref{coeffbased}.

The \blue{\emph{Markov state} of a receiver $r$ is} defined as \blue{the difference between the size of the transmission list and reception list,}
 \beq s_r(t)=|V_s(t)|-|V_r(t)|. \label{markovstate} \eeq
\blue{It should be noted that the removal of packets from the transmission queue does not affect the Markov state, since each packet removal decrements both $|V_s(t)|$ and $|V_r(t)|$.}

\subsection{The Markov chain model} \label{sec:markovmodel}
\begin{figure}\figtopsmall
\begin{center}
\begin{tikzpicture}[->,node distance=\imgsize/4.5]
\tikzstyle{every state}=[fill=none,draw=black,text=black,auto]
  \node[state] (A)                    {$0$};
  \node[state]         (B) [right of=A] {$1$};
  \node[state]         (C) [right of=B] {$2$};
  \node[state]         (D) [right of=C] {$3$};
  \node[state]         (E) [right of=D]       {$...$};

  \path (A) edge [loop above] node {$1-p$} (A)
            edge [bend left] node[above] {$p$} (B)
        (B) edge [loop above] node {$1-p-q$} (B)
            edge [bend left] node[below] {$q$} (A)
            edge [bend left] node[above] {$p$} (C)
        (C) edge [loop above] node {$1-p-q$} (C)
            edge [bend left] node [below] {$q$} (B)
            edge [bend left] node[above] {$p$} (D)
        (D) edge [loop above] node {$1-p-q$} (D)
            edge [bend left] node [below] {$q$} (C)
            edge [bend left] node[above] {$p$} (E)
        (E) edge [bend left] node [below] {$q$} (D);
\end{tikzpicture}
 \caption{A Markov chain describing \blue{transitions in the Markov state of a receiver $r$.}} \label{markov}
\end{center}
\figbottom
\end{figure}
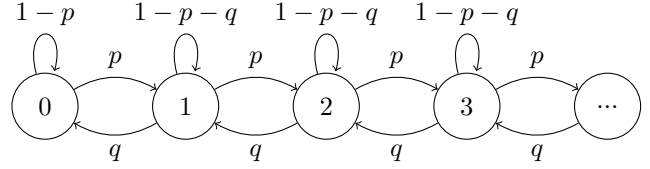

\blue{Under the baseline rate control scheme, if $\lambda<\mu$ then} changes to a receiver's Markov state over time can be modelled as a traversal through a Markov chain. This
is illustrated in Fig. \ref{markov}, where the states $0,1,2,...$
correspond to the values of $s_r(t)$. Whether $s_r(t)$ increases, decreases or
remains the same between time slots depends on both \blue{the add decision $a(t)$} and the receiver's channel conditions.
The allowable state transitions for states greater than zero and
their probabilities are listed in Table \ref{tab:markovtable}. Note
that \blue{as long as $\lambda<\mu$, the Markov chain is positive recurrent.} Although the Markov chain model is perfectly accurate for any
receiver considered on its own, the fact that the sender is shared
means that receivers' Markov states can exhibit a significant amount
of correlation with one another. Nevertheless this model still
provides valuable insight into the delivery delay characteristics of
the transmission schemes we will study.
\begin{table}
\center
\begin{tabular}{|c|c|c|}
 \hline
 State transition & Probability & Shorthand notation \\ \hline
$s_r(t+1)=s_r(t)+1$ & $\lambda \mubar$ & $p$ \\ \hline
$s_r(t+1)=s_r(t)-1$ & $\lambar \mu$ & $q$ \\ \hline
 $s_r(t+1)=s_r(t)$ & \blue{$\lambda \mu + \lambar \mubar$} & $1-p-q$  \\
\hline
\end{tabular}
\caption{The probability of transitions between Markov states for
$s_r(t)>0$, where the notation $\overline{x}=1-x$ is
used.}\label{tab:markovtable}
\figbottom
\end{table}

Using the concept of Markov state we can categorise the ways in
which the next packet $\pkt_n$ of a receiver $r$ can be delivered\footnote{This categorisation is also applicable to decoding without the in-order delivery constraint. Zero and leader state delivery both result in the decoding of all packets stored at the receiver, while \emph{coefficient-based decoding} results in the decoding of some subset of packets stored at the receiver.} to a receiver as follows.
\begin{enumerate}
\item {\bf \blue{Zero state delivery}}\\
\grn{Zero state delivery occurs when} a receiver $r$ is in the \emph{zero state}\grn{, i.e.} its Markov state
$s_r(t)=0$. At this point, the \blue{size of the reception list} equals the \blue{size of the transmission list}. Since coding schemes A, B and RLNC all satisfy the innovation
guarantee property, any time that $s_r(t)=0$, all packets \blue{in the transmission queue have been delivered}.
\item {\bf \blue{Leader state delivery}} \\
Under coding schemes A and B, a receiver $r$ is called a \emph{leader} if
it has \blue{the minimum} Markov state, i.e. $s_r(t)=\min{\{s_i(t)\}}$. \grn{Leader state delivery occurs when new packets are delivered by the current leader, although we require that }\blue{$s_i(t)>0$ to differentiate this from zero state delivery. In Section \ref{sec:coeff} we noted that the
effective \blue{transmission queue} is limited to the receiver with most packets
in their buffer. Therefore as shown in \cite{sundararajan1}, receiving a transmission while a leader results in the delivery of all packets \blue{in the effective transmission queue}.}
\item {\bf Coefficient-based delivery}\\
Under all three coding schemes, coefficient-based delivery accounts for any packets delivered to a receiver while it is neither leading nor in the zero state. \blue{Coefficient-based delivery occurs when the \grn{inclusion} of the transmission vector $v_s(t)$ \grn{into} the reception list of a receiver $r$ results in the decoding of the next needed packet $\pkt_n$.} In this case, some fraction of the packets \blue{stored at the receiver} are delivered.
\end{enumerate}

\subsection{Distribution of Markov states} \label{markovdistr}

Since the Markov state will form the basis of our analysis, the
first step is to find the probability $S_r(k)$ that at a randomly
selected time, the receiver $r$ is in state $k$. This is equivalent
to finding the stationary distribution of the Markov chain
corresponding to that receiver. For the Markov chain of Fig.
\ref{markov}, if the \blue{addition rate} $\lambda$ is less than the
channel rate $\mu$, a stationary distribution exists such
that
 \beq p S_r(k)=q S_r(k+1). \eeq
Solving for $\sum_{k=0}^{\infty}{S_r(k)}=1$, we obtain
 \beq S_r(k)=\left(1-\frac{p}{q}\right)\left(\frac{p}{q}\right)^k .
 \label{steadystate}\eeq

In the following Sections \ref{analysis} and \ref{coeffbased} we \med{shall} analyse the effect of Markov state on the receivers' delivery
delay.

\section{Zero and Leader State Delay Analysis} \label{analysis}

In this section we study the impact of the \blue{zero and leader state delivery} on the
receivers' delivery delay \blue{for the baseline transmission schemes outlined in Section \ref{sec:txschemes}}. By \grn{using the Markov state to distinguish} \blue{between different methods of packet delivery} we are able to provide insight
into the \red{delivery} behaviour of these throughput optimal \blue{coding} schemes that
has so far been missing from the literature. Taking zero state
delivery as a first approximation for our delay analysis, we use the
Markov chain model \blue{of Section \ref{sec:markovmodel}} to find the distribution of zero state delivery cycles, and accurately approximate the expected zero state delivery
delay. \blue{While} leader state delivery has proven an intractable
complication in previous analysis, we \grn{show how our model can be used to make useful observations about the impact of leader state delivery on the delivery delay.}

\subsection{Zero state delivery} \label{sec:zdec}\label{deccycle}

\begin{figure}\figtopsmall
\begin{center}
\includegraphics[width=\imgsize]{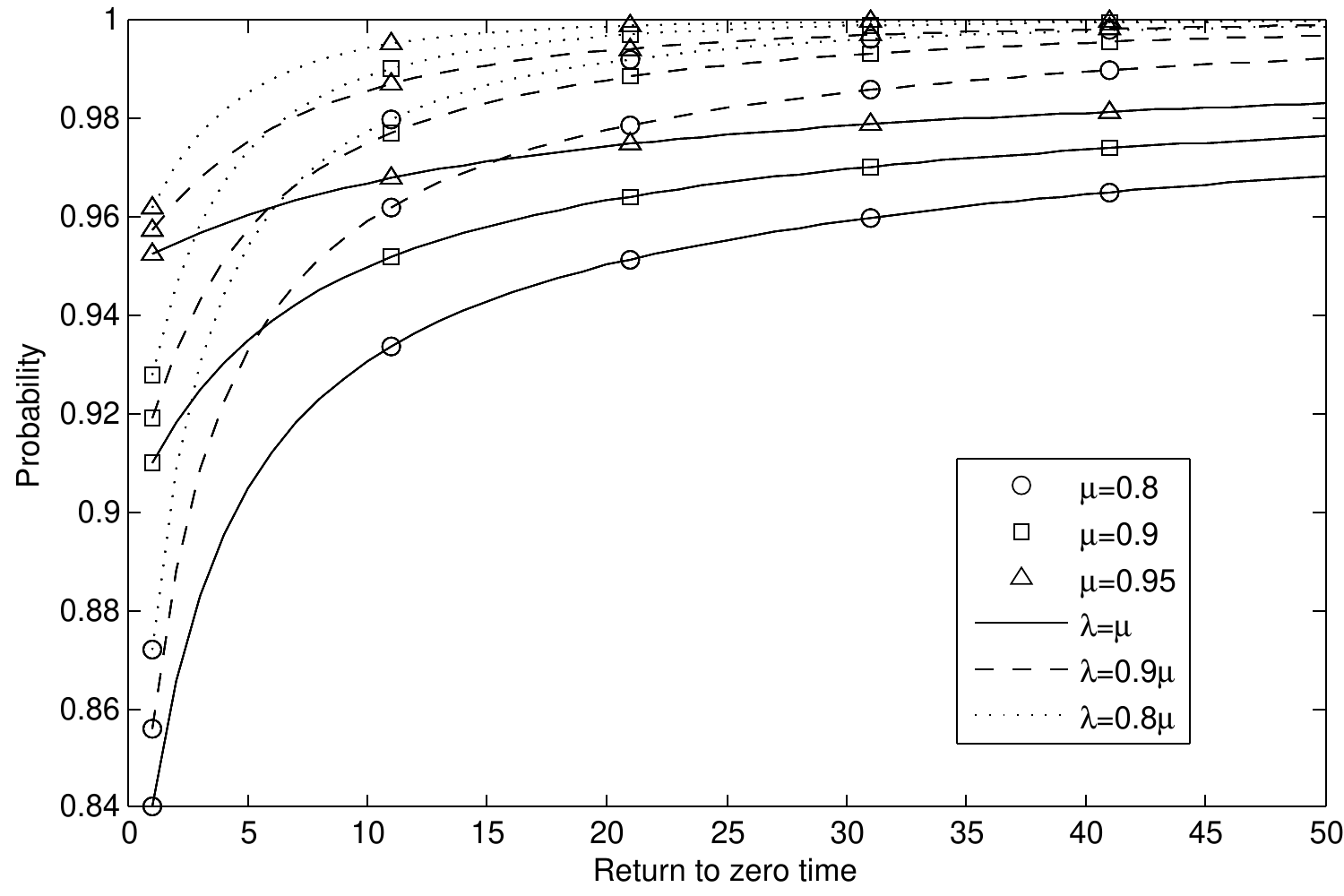}
\caption{The probability from \eqref{zerotozero} of a receiver
taking $\leq T$ time steps to return to the zero state \red{under the baseline rate control scheme}.}
\label{zerocycle}
\end{center}
\figbottom
\end{figure}

Here we will \blue{estimate the \emph{zero state delivery delay}, defined as the \blue{delivery delay experienced if only zero state delivery is permitted}. This estimate will be used} as an upper bound on the delivery delay for \blue{the baseline transmission schemes}. It is important to observe that, as long as the innovation guarantee property holds, the Markov state of a receiver depends only on its \blue{channel} rate \grn{$\mu$} and the \blue{addition rate} \grn{$\lambda$}. \blue{Therefore zero state delivery} is not affected by the coding scheme, the presence of other receivers, or
even the quality of feedback. This independence makes zero state
delivery analysis a valuable tool, as
initial performance estimates \blue{can be made} without the intractable complications
that have hindered the study of \blue{network coded} transmission schemes to date.

To find the zero state \blue{delivery} delay, it is not sufficient to know
the proportion of time a receiver spends in the zero state,
calculated in \eqref{steadystate}. The \blue{zero state delivery delay} depends
on the distribution of times between returns to the zero state,
which we call \emph{\red{delivery} cycles}\grn{, and the distribution of transmission queue additions within each cycle}. Therefore, we \med{shall} use random
walk analysis to calculate the distribution of \red{delivery} cycle
lengths, and based on this work, find an accurate approximation for
the zero state \blue{delivery} delay of \blue{\red{baseline} transmission schemes}.

\subsubsection{\red{Delivery} cycle distributions} 
A receiver starting in Markov state 0 experiences a \emph{\red{delivery}
cycle} of length $T$ if its first return to the zero state in the
Markov chain occurs after exactly $T$ time slots. We calculate
$P_{0,0}(T)$, the probability that a \red{delivery} cycle will be of
length $T$.

We can solve this problem in two steps. First, we characterise a
path through the Markov chain that consists of only \emph{moving
steps} where $s_r(t+1)=s_r(t)\pm 1$. Then we factor in the effect of
\emph{pause steps}, where $s_r(t+1)=s_r(t)$.

In the first time step there are two possibilities. The receiver can
remain at \blue{state} 0 with probability $ 1-p$, which gives us
$P_{0,0}(1)=1-p$. If it instead moves up to state 1, it must return
to 0 in $T>1$ time steps. For a path of fixed length $T$ to start at
and return to 0, it must consist of $2k$ moving steps, $k$ up and
$k$ down, and $T-2k$ pause steps, where $1\leq k \leq \lfloor T/2
\rfloor$. If no other encounters with the zero state are permitted,
the first and last time steps must be up and down steps
respectively. Therefore the number of paths that first return to the
zero state in exactly $2k$ steps without pauses is given by the
$(k-1)$-th Catalan number \cite{brualdi99}
 \beq C_{k-1}=\frac{1}{k} {2k-2 \choose k-1}.
 \eeq
Now we factor in the $T-2k$ pauses. These pauses cannot occur in the
first or last time step, otherwise the \red{delivery} cycle length would
not be $T$. For a given path of $2k$ moving steps, there are ${T-2
\choose 2k-2}$ choices for pause locations. Therefore the
probability of taking exactly $T>1$ timeslots to return to the
\blue{zero state} is given by
 \beq P_{0,0}(T)= \sum_{k=1}^{\lfloor T/2 \rfloor} \frac{1}{k} {2k-2 \choose k-1}
 {T-2 \choose 2k-2} p^{k}q^{k}(1-p-q)^{T-2k} . \label{zerotozero}
 \eeq
The cumulative \red{delivery} cycle length probabilities are given for a
number of \blue{values of} $\lambda$ and $\mu$ in Fig.
\ref{zerocycle}. The greater the load factor $\rho$, the more slowly
the probability converges to 1 and the larger the \blue{zero state delivery delay}.

\subsubsection{Delay estimate} \label{sec:zdelayest}
\red{Over a delivery cycle of length $T$, we estimate of the number of packets added to the transmission queue and their expected delivery delay. This is combined with \eqref{zerotozero} to obtain an accurate estimate of the zero state delivery delay.}

Where the \red{delivery} cycle is of length $T=1$, the probability that one
packet \blue{is added and then} immediately \red{delivered} is simply $\lambda\mu$.
\blue{Since the packet is immediately \red{delivered}, it incurs no} \red{delivery} delay.

In Section \ref{deccycle} we established that, for all other coding
cycles with length $T\geq 2$, the Markov state must increase in the
first time slot, and decrease in the last time slot. Therefore, \blue{we must have $a(t)=1$ in the first time slot, and $a(t)=0$ in the} last time slot. We now \grn{assume} that in the remaining
$T-2$ time slots \blue{additions occur} uniformly with probability
$\lambda$.

\grn{Then} the average number of packets delivered over a \red{delivery} cycle of
length $T$ is estimated to be
 \beq 1+\lambda (T-2) \eeq
\grn{and} the \blue{average zero state delivery delay for} \grn{each of} these packets is $T/2$.
The \red{total} delay incurred by these packets \grn{would then be}
 \beq T+ 0.5 \lambda T(T-2). \eeq

Therefore the \blue{zero state delivery delay}, including the $T=1$ \red{delivery} cycle, \grn{can be estimated as}
 \beq \frac{\sum_{T=2}^{\infty} P_{0,0}(T) (T+0.5 \lambda T(T-2))}
 {\lambda\mu+1+ \sum_{T=2}^{\infty}P_{0,0}(T) \lambda (T-2)}. \label{zdeccalc} \eeq

\begin{figure}\figtopsmall\begin{center}
\includegraphics[width=\imgsize]{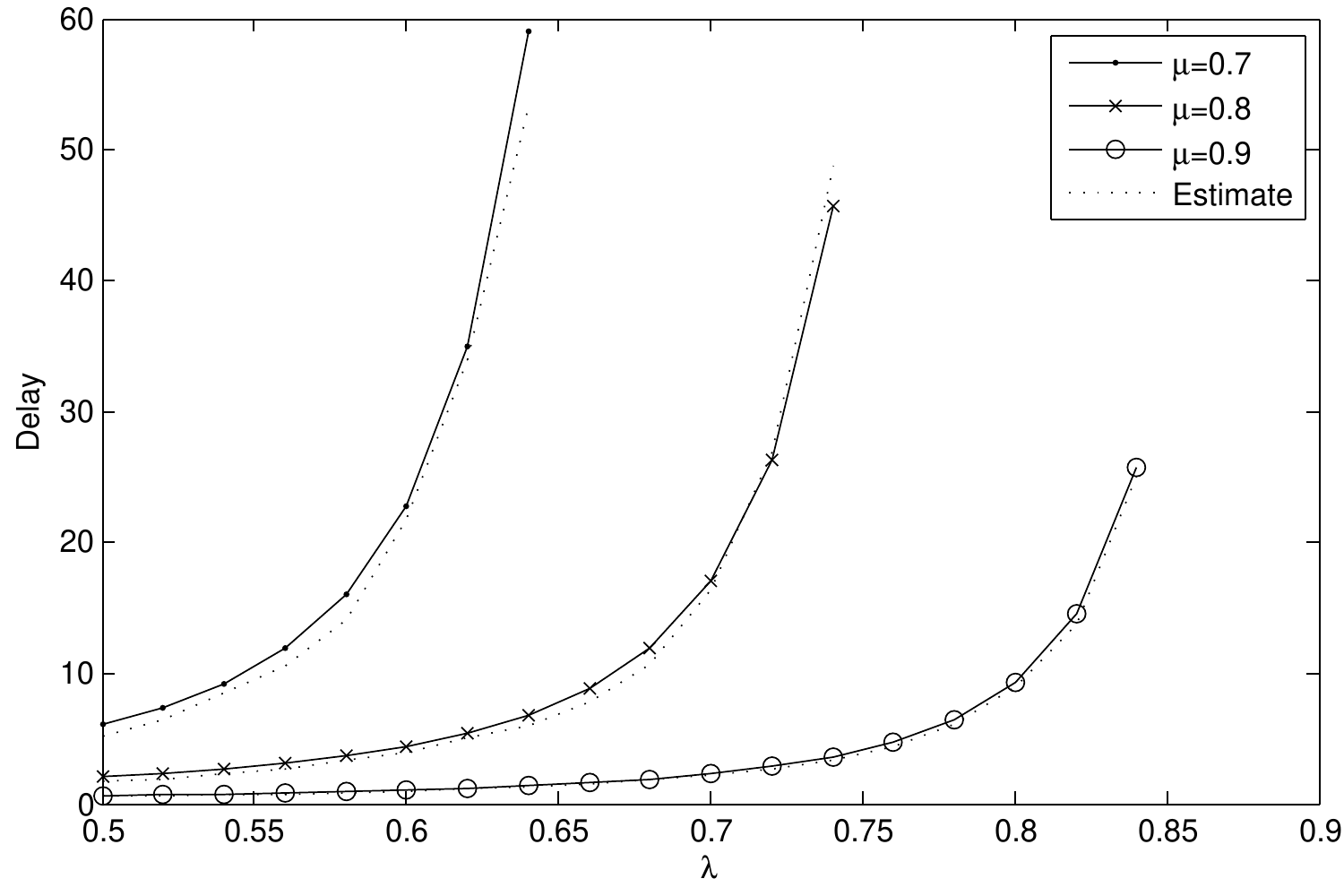}
\caption{The \blue{zero state delivery delay of the baseline rate control scheme}, as a function of the addition rate $\lambda$. \blue{The \grn{delay estimates} of} \eqref{zdeccalc}
(dotted lines) \blue{are compared against} simulation (solid lines).}
\label{zdecest}
\end{center}\figbottom
\end{figure}

In Fig. \ref{zdecest} we show how our calculated estimate, truncated
at $T=1000$, matches well with the average delivery delays obtained
from simulation.

\subsection{Leader state delivery}

\blue{In this section we study the \emph{leader state delivery delay}, defined as the delivery delay experienced if only zero and leader state delivery are allowed.} \red{Note that as mentioned in Section \ref{sec:coeff}, leader state delivery has an equal impact on coding schemes A and B, but does not apply to the RLNC scheme.} We investigate the amount of time receivers spend leading and its impact on the delivery delay, \blue{compared with} zero state delivery \blue{on its own}.

\subsubsection{Leader state distribution}
\begin{figure}\figtopsmall
\begin{center}

\begin{tikzpicture}[->,node distance=\imgsize/5, bend angle=10]
\tikzstyle{every state}=[fill=none,draw=black,text=black,auto]
  \node         (A0)                {};
  \node         (B0) [right of=A0] {};
  \node         (C0) [right of=B0] {};

  \node         (A1) [above of=A0] {};
  \node[state]  (B1) [right of=A1] {\small{$i,j$}};
  \node         (C1) [right of=B1] {};

  \node         (A2) [above of=A1] {};
  \node         (B2) [right of=A2] {};
  \node         (C2) [right of=B2] {};

  \path
    (B1) edge node[auto] {\small{$\lambar \mu^2$}} (A0)
    (B1) edge node[auto] {\small{$\lambar \mu \mubar$}} (A1)
    (B1) edge node[auto] {\small{$\lambar \mubar \mu$}} (B0)
    (B1) edge node[auto] {\small{$\lambda \mu \mubar$}} (B2)
    (B1) edge node[auto] {\small{$\lambda \mubar \mu$}} (C1)
    (B1) edge node[auto] {\small{$\lambda \mubar^2$}} (C2)
  ;
\end{tikzpicture}
\begin{tikzpicture}[->,node distance=\imgsize/5, bend angle=10]
\tikzstyle{every state}=[fill=none,draw=black,text=black,auto]
  \node (C0) {};
  \node (C1)[above of= C0] {};
  \node (C2) [above of= C1] {};
  \node (D0) [right of=C0] {};
  \node[state] (D1) [above of=D0] {\small{$i,j$}};
  \node (D2) [above of=D1] {};
  \node (E0) [right of=D0] {};
  \node (E1) [above of=E0] {};
  \node (E2) [above of=E1] {};

  \path
    (D1) edge node[auto] {\small{$q^2$}} (C0)
    (D1) edge node[right] {\small{\begin{sideways}$(1\mbox{-}p\mbox{-}q)q$\end{sideways}}} (D0)
    (D1) edge node[auto] {\small{$pq$}} (E0)
    (D1) edge node[auto] {\small{$q(1\mbox{-}p\mbox{-}q)$}} (C1)
    (D1) edge node[auto] {\small{$p(1\mbox{-}p\mbox{-}q)$}} (E1)
    (D1) edge node[auto] {\small{$qp$}} (C2)
    (D1) edge node[left] {\small{\begin{sideways}$(1\mbox{-}p\mbox{-}q)p$\end{sideways}}} (D2)
    (D1) edge node[auto] {\small{$p^2$}} (E2)
  ;
\end{tikzpicture}
\caption{Two-receiver state transition probabilities (a) in
practice, and (b) under the independent receiver model. The
horizontal and vertical axes correspond to the Markov states of
receivers 1 and 2 respectively, where $s_1(t)=i$ and $s_2(t)=j$.
Since the sender is common to both receivers, it is not possible
that $s_1(t+1)=i\pm 1$ while $s_2(t+1)=j\mp 1$.}\label{tworxindep}

\end{center}\figbottom
\end{figure}
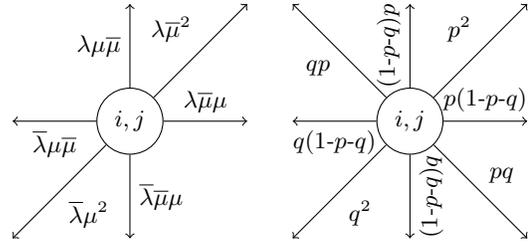

\begin{figure}\figtopsmall
\begin{center}
\includegraphics[width=\imgsize]{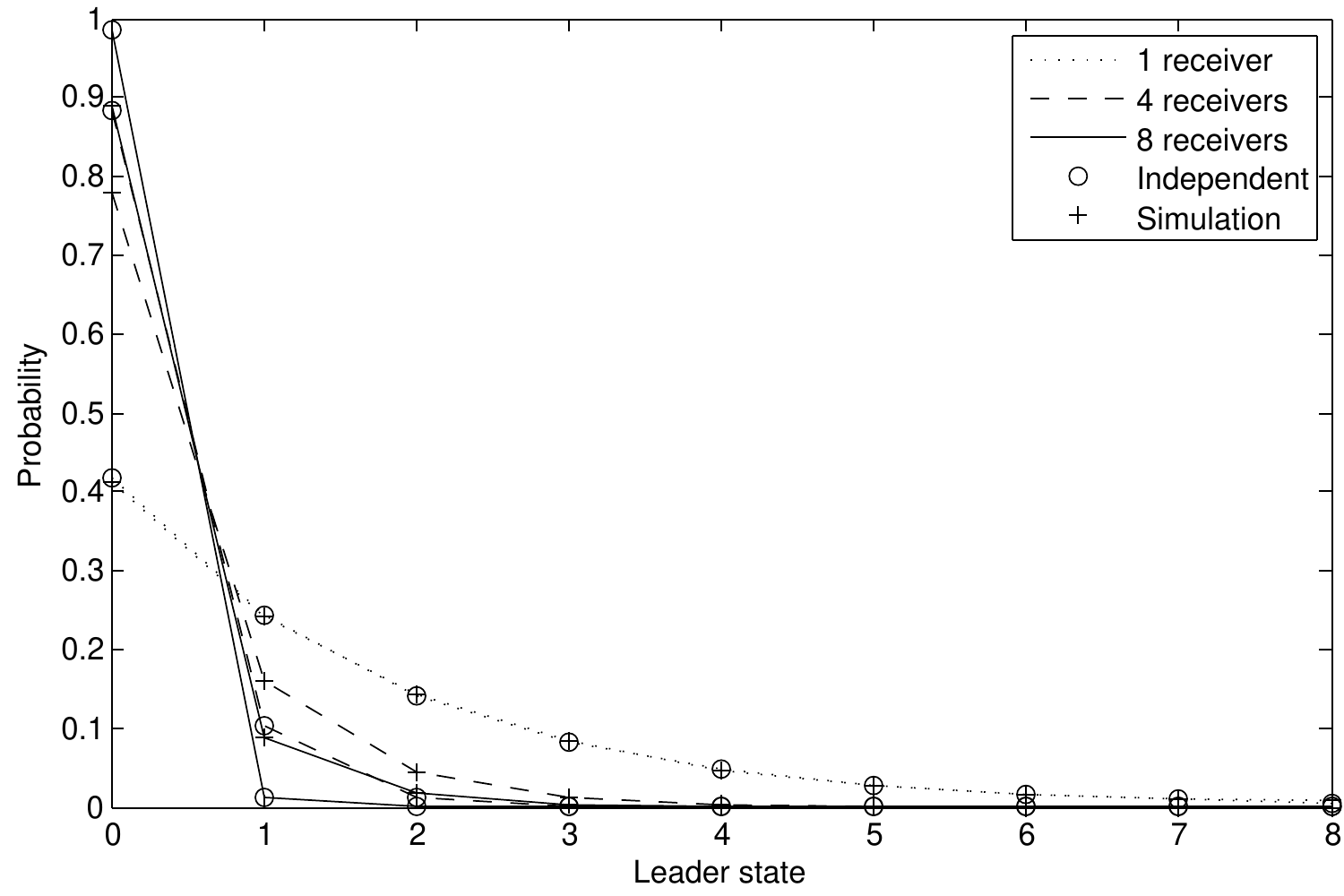}
\caption{\blue{Under the baseline rate control scheme with addition rate $\lambda$,} the proportion of time the leader is in each state, under
the independent receiver model \eqref{leadindep} and in practice.
$\lambda=0.7,\mu=0.8$.} \label{leading}
\end{center}\figbottom
\end{figure}

Based on the Markov state distribution calculated in \ref{markovdistr}, we bound the
average time that the leader(s) spend in each Markov state.
By \eqref{steadystate}, the probability of a receiver $r$ being in a
state $\geq k$ is
 \beq S_r(\geq k)=\sum_{i=k}^{\infty} S_r(i)
 =\left(\frac{p}{q}\right)^k. \eeq
\grn{So if} receivers' Markov states \grn{were} independent, the probability of having
a leader in state $k$ would be
 \beq L(k)=\left(1-\left(\frac{p}{q} \right)^R \right) \left( \frac{p}{q} \right)
 ^{Rk}. \label{leadindep}
 \eeq
However, since the sender is common to all receivers, there is a
noticeable amount of correlation between receivers' Markov states.
This is illustrated in Fig. \ref{tworxindep}, which compares the
joint Markov state transition probabilities for two receivers under
each model. In practice the correlated transition probabilities
result in the receivers being more closely grouped together than
predicted by the independent receiver model. Fig. \ref{leading}
shows that the probability of leading from states $k>0$ is higher in
practice than under the independent receiver model in
\eqref{leadindep}.

\subsubsection{Observations} \label{sec:leaderobs}
The probability of the leader being in state $k$ is bounded between
the values in the single receiver case and the independent receiver
model. Therefore, although the independent receiver model is not
entirely accurate, it can still be used to make \blue{the following}
observations about the leader state.
\begin{enumerate}
 \item The probability that a receiver $r$ is leading is
$\geq 1/R$, since at least one receiver must lead at each time slot.
 \item The leader will \blue{most likely} be in state $k=0$. The larger \blue{the number of receivers} $R$, the more likely this is the case.
 \item The higher a receiver's state $k$, the lower its likelihood of leading.
 \item By \eqref{steadystate} and \eqref{leadindep} as the load factor $\rho
 \rightarrow 1^-$, or equivalently
$\lambda \rightarrow \mu^-$, the state probability distribution
$S_r(\geq k)$ converges on 1 more slowly. This increases the
probability that the leader will be in \blue{a state $k>0$}, and therefore
the impact leader state delivery has on delay.
\end{enumerate}

\begin{figure}\figtopsmall\begin{center}
\includegraphics[width=\imgsize]{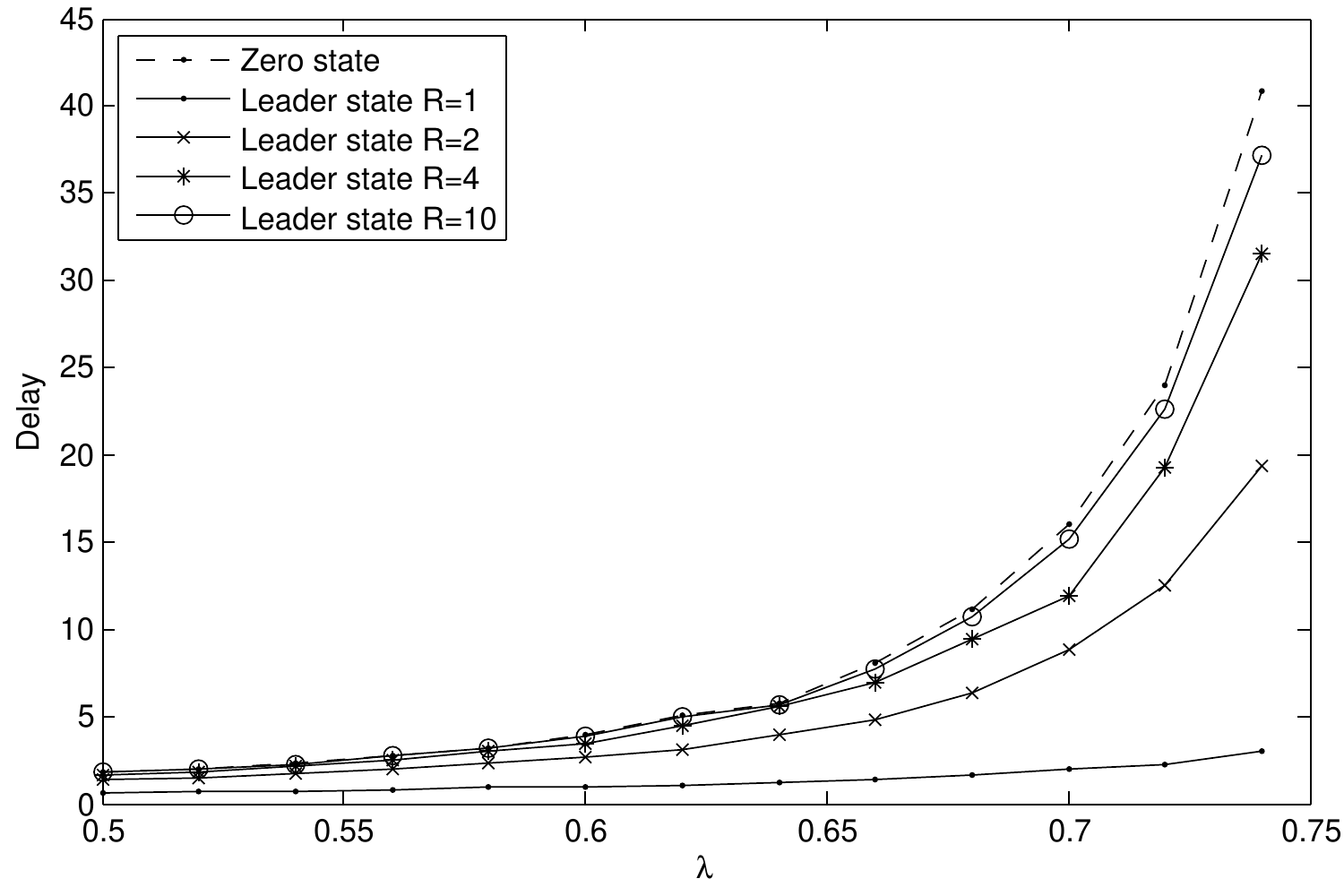}
\caption{\blue{Simulated zero state and leader state \blue{delivery delays under the baseline rate control scheme with addition rate $\lambda$},
for different numbers of receivers $R$.} $\mu=0.8$.} \label{rxleader}
\end{center}\figbottom\end{figure}

We can observe some of these effects in Fig. \ref{rxleader}. $R=1$ represents the extreme case where \blue{there is only one receiver who} is always leading, and so results in extremely low delivery delays. As $R$ increases, however, the \blue{leader state delivery} delay quickly
\blue{converges towards the} zero state delivery \blue{delay}. Even at moderate values of
$R$, for example $R=10$, the \blue{difference between the zero and leader state delivery delay}
is negligibly small. \blue{This behaviour can be attributed to observations 1 and 2, made above.} \blue{By contrast,} as the load factor \blue{$\rho$} increases, so does the \blue{impact of} leader state delivery, consistent with observation 4.

Under imperfect feedback conditions, \blue{the contribution of leader state delivery would be further diminished}, since \blue{the sender would not always know which} packets the leader has received. In order to maintain the innovation guarantee property, the sender \blue{would need to account for the possibility the leader has received all packets for which the outcome has not yet been determined. This overestimation of the leader's channel rate would result in an effective transmission queue closer in size to \red{that of the actual} transmission queue.}

\section{Coefficient-based Delivery}\label{coeffbased}

Coefficient-based delivery accounts for \blue{any} remaining packets \blue{delivered while a receiver is neither leading nor in the zero state}. The impact of coefficient-based delivery is not well understood because of the difficulty of analysing its effects. In the literature it is
generally speculated to contribute a small, if not negligible,
improvement on the delivey delay. However through simulation we
demonstrate two important principles for improving the likelihood of
coefficient-based delivery: minimising the coding field size $M$, and
maintaining sparse codes \blue{(i.e. minimising the number of nonzero \red{coefficients $\alpha_i(t)$ in \eqref{eq:cod}})}. When these conditions are met,
coefficient-based delivery can reduce the delivery delay
significantly.

Say that at time $t$ the sender transmits a packet $c(t)$ with
\red{transmission vector} $v_s(t)$. Then \blue{using the concepts from Section \ref{sec:kspace},} the next needed packet will be delivered \red{if and}
only if the following condition holds.
\begin{lemma} \label{chdeclemma}
At time $t$, a receiver can \grn{deliver} \blue{their} next needed packet $\pkt_n$
iff \red{they receive a packet with transmission vector} $v_s(t) \in \spn(K_r(t-1) \cup e_n)
\setminus K_r(t-1)$.
\end{lemma}
\begin{IEEEproof}
A packet $\pkt_k$ is decoded iff $e_k \in K_r(t)$. Say that $e_n
\notin K_r(t-1)$. Then for $\pkt_n$ to be \grn{delivered} at time $t$, $e_n
\in \spn(K_r(t-1) \cup v_s(t))$. To satisfy the innovation guarantee
property, $v_s(t) \notin K_r(t-1)$. Therefore to \grn{deliver} packet
$\pkt_n$ at time $t$, $v_s(t) \in \spn(K_r(t-1) \cup e_n) \setminus
K_r(t-1)$.
\end{IEEEproof}

As we \med{shall} show, the probability of coefficient-based delivery
depends on both the \blue{coding} scheme used and the \emph{effective
Markov state}, which we now define.

\subsection{Effective Markov state} \label{sec:effmarkov}

The \emph{effective transmission list} $V^*_s(t)$ is defined as
the set of \blue{basis} vectors corresponding to \blue{packets in the effective transmission queue}. In the RLNC scheme, typically $V^*_s(t)=V_s(t)$, unless all coefficients selected for \blue{the \grn{newest}} packet happen to be 0.
In contrast, under coding schemes A and B \blue{the effective transmission queue is limited by the \grn{the number of packets known by} the leading receiver(s), so that}
 \beq \blue{|V^*_s(t)|}=\min(\max_{r \in 1,...,R}(|V_r(t-1)|+1),|V_s(t)|) \eeq
\blue{and} $V^*_s(t)=\{e_1,e_2...,e_{|V^*_s(t)|}\}$. \med{Similarly} to
\eqref{eq:kspace}, the \emph{effective \blue{transmission} knowledge space} is
$K^*_s(t)=\spn(V^*_s(t))$. The \emph{effective Markov state} of a
receiver $r$ can then be defined as
 \beq s^*_r(t)= |V^*_s(t)|-|V_r(t-1)|. \label{effmarkov}\eeq
This differs from \eqref{markovstate} in that, in order to calculate
the probability that the \emph{current} transmission $c(t)$ will \red{deliver $\pkt_n$}, it
compares the effective \blue{transmission list} to the reception list
\emph{prior} to \blue{packet receptions in} the current time slot.

\begin{lemma}\label{lem:2}
A receiver $r$ can only coefficient-based \red{deliver} its next needed
packet $\pkt_n$ when its effective Markov state decreases, i.e.
$s^*_r(t)=s^*_r(t-1)-1$.
\end{lemma}
\begin{IEEEproof}
It is always true that $K_r(t-1) \subset K^*_s(t-1)$.
If the receiver is not a leader, then they have not decoded all
packets in $K^*_s(t-1)$ and $e_n \in K^*_s(t-1)$. Therefore by Lemma
\ref{chdeclemma}, coefficient-based delivery can only occur if
$v_s(t) \in K^*_s(t-1)$, so that $V^*_s(t)=V^*_s(t-1)$. Receiving an
innovative packet means that $|V_r(t)|=|V_r(t-1)|+1$, so by
(\ref{effmarkov}) $s^*_r(t)=s^*_r(t-1)-1$.
\end{IEEEproof}

So in order for a coefficient-based delivery opportunity to arise,
three conditions must \blue{first} be satisfied:
\begin{itemize}
 \item The receiver is neither a leader nor in the zero state
 \item No new packets are encoded by the sender
 \item The receiver successfully receives the transmitted packet.
\end{itemize}

Therefore, of the time slots a receiver is neither in the zero state
or a leader, approximately $\lambar \mu$ of these provide \blue{an}
opportunity for coefficient-based delivery to occur. We now
investigate the effectiveness of \blue{coding} schemes A
and B and RLNC in utilising this fraction of
\emph{coefficient-based \red{deliverable}} time slots to minimise delay.

\subsection{RLNC scheme}
To gain some insight into the probability of coefficient-based
delivery, we first study the RLNC scheme. Here we will
demonstrate how the effective Markov state affects the probability
of coefficient-based delivery.

We first calculate for a single receiver the probability that with
\blue{receiver} knowledge space $K_r(t)$, the next needed packet $\pkt_n$ will be
delivered. The total number of possible transmissions is given by
the size of the \blue{transmission} knowledge space $K_s(t)$ minus the
receiver \grn{knowledge space}. Therefore by Lemma \ref{chdeclemma}, the probability of
selecting a packet under the RLNC scheme which allows $\pkt_n$
to be delivered is
\begin{align}
& \frac{|\spn(K_r(t-1) \cup e_n) \setminus K_r(t-1)|}{|\blue{K^*_s(t)} \setminus K_r(t-1)|} \nonumber \\
&= \frac{M-1}{M^{s^*_r(t)}-1} \label{baselinedec}
\end{align}
Therefore, the probability of coefficient-based delivery depends
only on the \blue{effective} Markov state of the receiver and the field size $M$. The
exponential dependence on both of these factors means that the
coefficient-based delivery probability will be very small for high
effective Markov states and large field sizes.

For the multiple receiver case, simulations show that
there is a \grn{fairly} negligible difference between the RLNC
coefficient-based delivery probabilities for the single
and multiple receiver cases, provided they are coded using
the same field size \blue{$M$}. Some of these
probabilities, normalised over coefficient-based \red{deliverable}
time slots, are shown in Fig. \ref{chdec}, and the
resulting delay performance for $M=4$ is given in Fig.
\ref{allchdec}. As expected, the small coefficient-based
delivery probability results in only a slight improvement
over zero state delivery.

\subsection{Coding scheme A}
Under coding scheme A, the sender codes only the first unseen packet of
each receiver. Furthermore the coefficient\grn{s} chosen \grn{are} the smallest
that will satisfy the innovation guarantee property.\footnote{In
\cite{sundararajan3} it is suggested that any coefficient satisfying
the innovation guarantee is suitable, but in our implementation the
smallest allowable coefficient is chosen.} Although a field size $M
\geq R$ is necessary to guarantee innovation, the majority of the
time coefficients from a much smaller field size $\mathbb{F}_{2}$
are sufficient.

In Fig. \ref{chdec} the coefficient-based delivery \grn{probability of}
coding scheme A is compared against the RLNC scheme. \grn{Under} coding scheme A with four receivers and $\mathbb{F}_4$, \grn{the probability of coefficient-based delivery} in fact \grn{lies between the probabilities for the $\mathbb{F}_2$ and $\mathbb{F}_4$} RLNC scheme. This occurs since \blue{in practice the sender usually selects binary field coefficients, effectively coding from the field $\mathbb{F}_2$}. With 8
receivers and a field $\mathbb{F}_8$, \grn{packet coefficients are nearly always selected from the field $\mathbb{F}_{4}$}. This
results in \blue{coefficient-based delivery} probabilities \grn{close} to the RLNC
$\mathbb{F}_4$ case. In Fig. \ref{allchdec} we can observe that
coding scheme A has significantly better delay performance compared with
the RLNC scheme. This is primarily due to the role of leader
state delivery, with a slight contribution from coefficient-based
delivery.

\subsection{Coding scheme B}

Coding scheme B, by contrast, attempts to closely mimic a systematic,
uncoded scheme by coding additional packets into each transmission
only if it is necessary to maintain the innovation guarantee \blue{property}. Each
of these extra packets has the additional property that, if
received, the coded transmission will allow the corresponding
receiver to deliver their next needed packet.

\begin{figure}\figtopsmall
\begin{center}
\includegraphics[width=\imgsize]{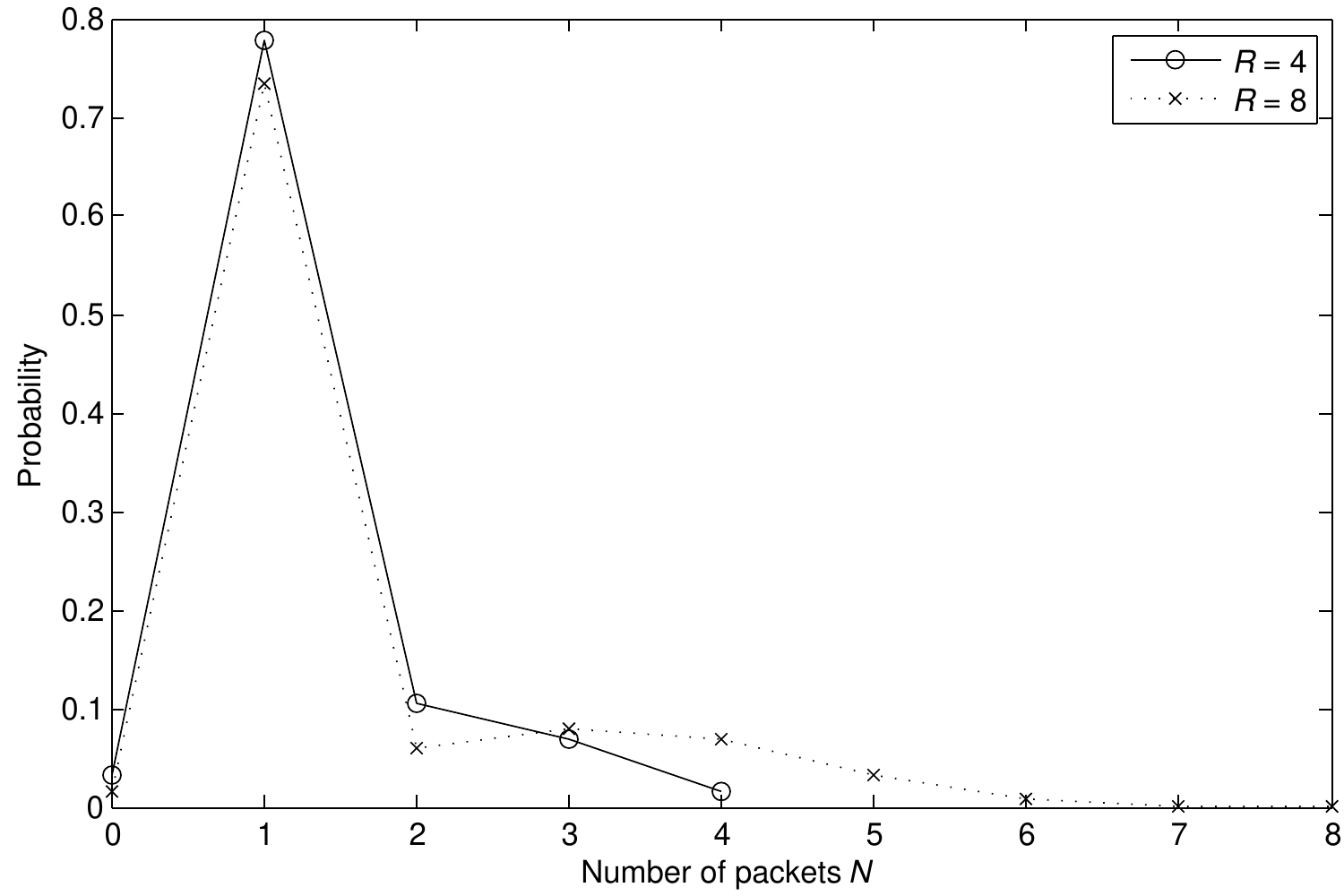}
\caption{For the baseline rate control scheme with coding scheme B, the \blue{probability that $N$} packets are coded \blue{into a} sender transmission. $R=4,8$, $\lambda=0.7$, $\mu=0.8$. Note that, \blue{if the transmission queue is empty, no packets will be} coded.}\label{s2senderpkts}
\end{center}\figbottom
\end{figure}

We can expect that at least $\lambda$ of the sender's transmissions,
corresponding to the first transmission of each new packet, will be
uncoded. Fig. \ref{s2senderpkts} agrees with this prediction, and we
can observe that the four-receiver case has a \blue{slightly} higher
proportion of uncoded packets compared with the eight receiver case.
This can be attributed to the fact that the smaller the number of
receivers, the lower the probability that additional packets need to
be included in each transmission.

%

\begin{figure}\figtopsmall
\begin{center}
\includegraphics[width=\imgsize]{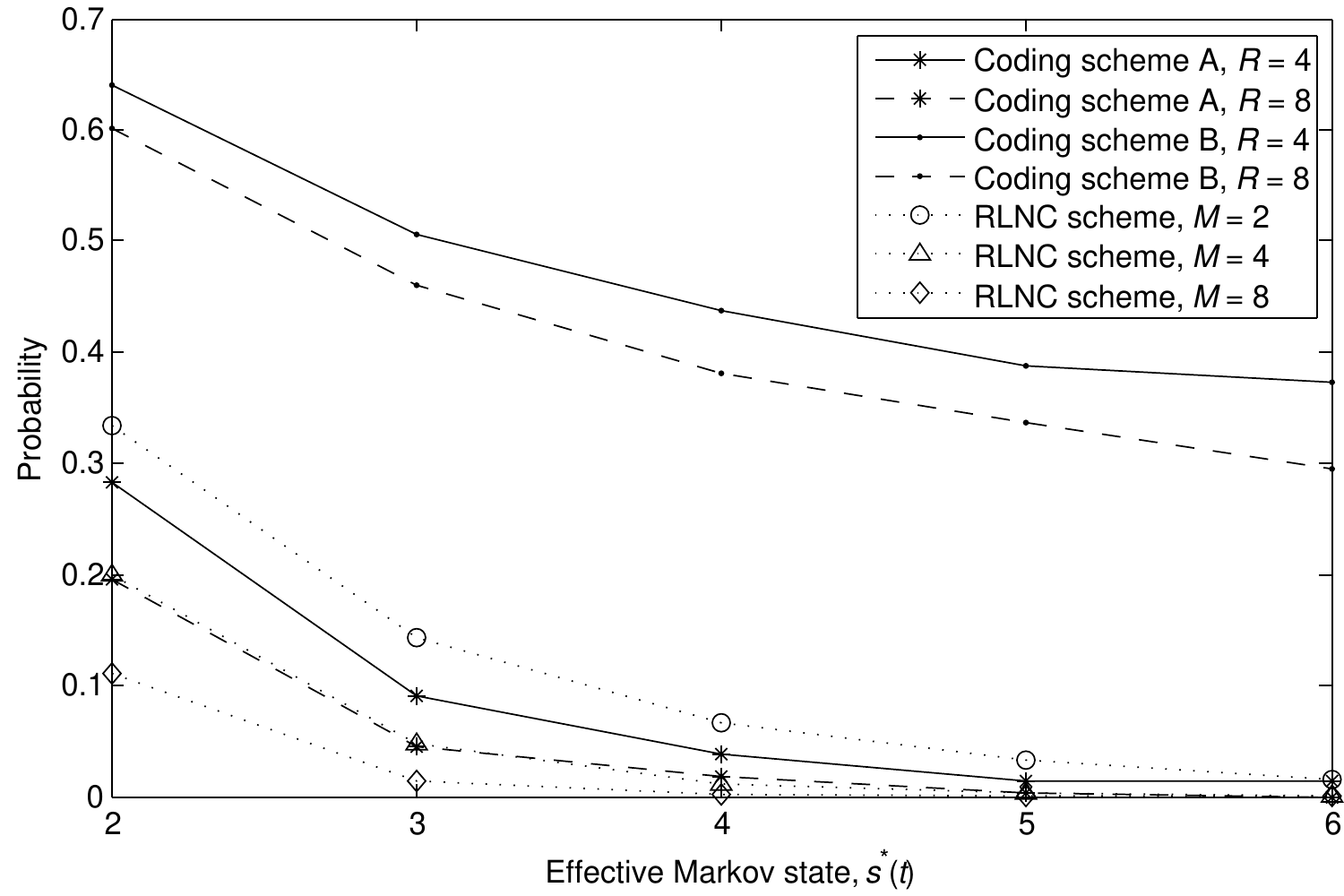}
\caption{\blue{Probability of coefficient-based delivery as a function of Markov state, for baseline transmission schemes with addition rate $\lambda=0.7$.} Probabilities are normalised over coefficient-based \red{deliverable} timeslots. The field size $M=R$ the number of receivers, and $\mu=0.8$.} \label{chdec}
\end{center}\figbottom
\end{figure}

\begin{figure}\figtopsmall
\begin{center}
\includegraphics[width=\imgsize]{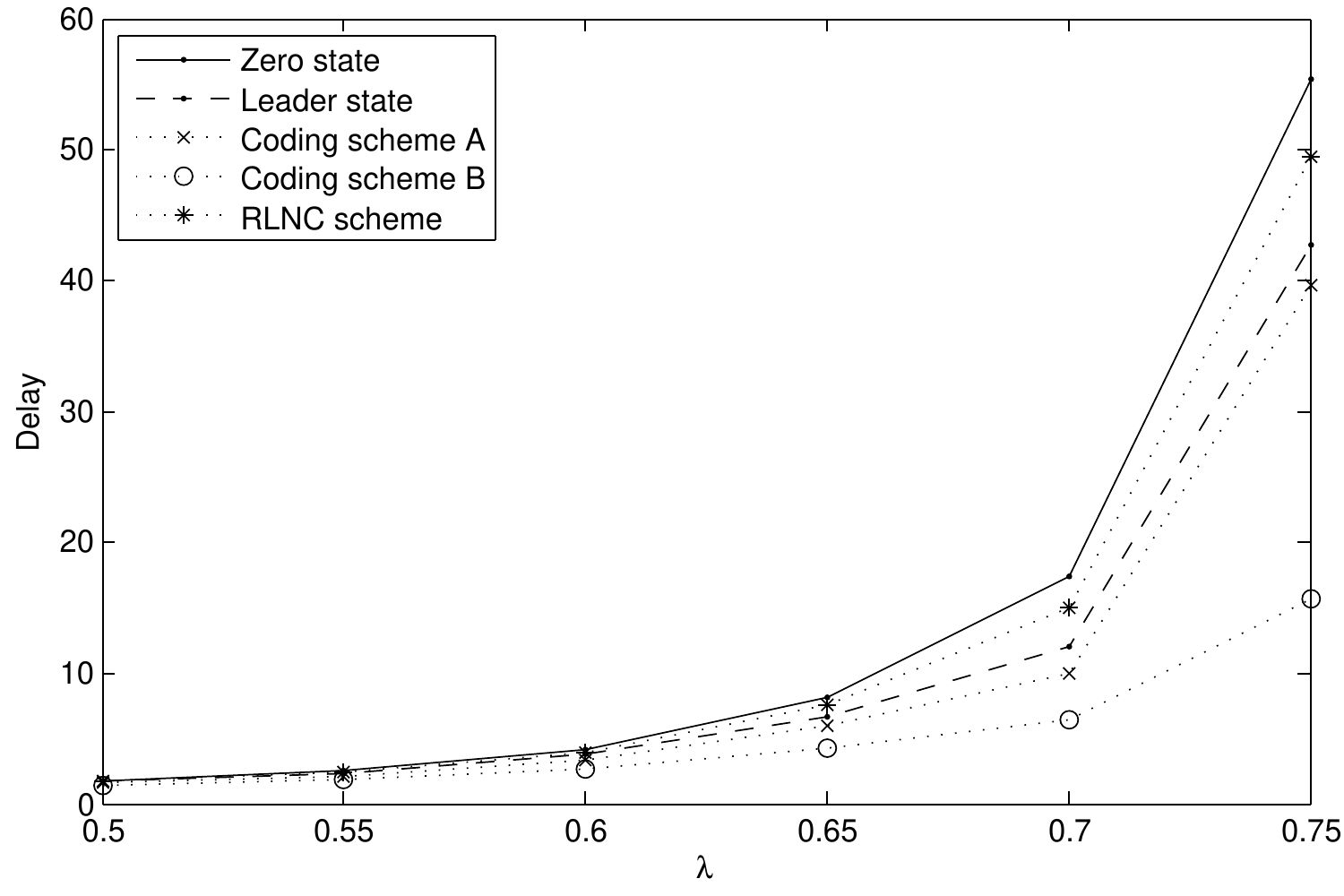}
\caption{The average delivery delay of \blue{coding} schemes A, B and \blue{RLNC
under the baseline rate control scheme with addition rate $\lambda$}. Zero and leader state
\blue{delivery delays} are included for comparison.
$R=4,\mu=0.8$.}\label{allchdec}
\end{center}\figbottom
\end{figure}

Coding scheme B is shown in Fig. \ref{chdec} to have a coefficient-based
delivery probability that is significantly higher than coding scheme A and
decays more slowly as a function of the effective Markov state. The \blue{coefficient-based delivery} probability for the eight-receiver case, which has a smaller fraction of uncoded packets, is somewhat less than its four-receiver counterpart. From Fig. \ref{allchdec} we can observe that the higher coefficient-based delivery probabilities of coding scheme B
result in significantly better delivery delay compared with both
coding scheme A and the RLNC scheme. The improvements are especially
notable at high \red{addition} rates, with an almost threefold
improvement in \blue{the} delivery delay compared with \blue{the leader state delivery delay}.

We can give an intuitive explanation for the link between \red{coding} sparsity
and a higher probability of coefficient-based delivery. If a large
fraction of undelivered packets are already decoded, this
effectively reduces the size of the system of equations
corresponding to unknowns in the receiver's buffer. If the
transmitted combination is itself sparse, then there is a good
probability that its few nonzero elements are those previously
decoded by the receiver. Where the elements corresponding to other
receivers are already known, the sender will ensure that the
transmitted combination allows the delivery of the receiver's next
needed packet.

It should be noted that under \blue{coding schemes A and B},
\red{infrequent} feedback \red{could potentially} degrade the delivery delay performance. If
the sender were to make decisions based on incomplete information
about the contents of receivers' buffers, \blue{throughput optimality would only be achievable at the cost of larger field sizes and less sparse coding, both of which would have a detrimental impact on the delivery delay.}

\section{Dynamic rate control scheme} \label{dynamicscheme}

\bII{In this section we outline the decision metric used in the \emph{dynamic rate control scheme}. This rate control scheme, which builds upon the Markov state model analysed in Sections \ref{sec:model} to \ref{coeffbased}, determines whether the sender adds new packets to the transmission queue or waits, based on the receivers' predicted throughput and delay performance. Using this decision metric, we will demonstrate in Section \ref{sec:performance} that improved throughput-delay performance can be achieved, compared with both the baseline and delay threshold rate control schemes.

\subsection{Dynamic delivery model}
The first step is to outline the \emph{dynamic delivery model} used by the dynamic rate control scheme to predict the receivers' throughput and delivery performance. Under the dynamic delivery model, we continue to model sender additions to the transmission queue by a Bernoulli process, as was done for the baseline rate control scheme. Although the dynamic rate control scheme clearly differs from the baseline rate control scheme, the dynamic nature of the rate control scheme means that there is no straightforward way to predict when the sender will choose to add. Therefore, the zero state delivery model continues to model sender additions with a Bernoulli process, replacing the previously known addition rate $\lambda$ with an \emph{addition rate estimate}, $\lambde$. Using \eqref{eq:totarriv}, $\lambde$ is given by the average observed addition rate, upper bounded by $\mu$, the maximum possible delivery rate,
 \beq \lambde= \min(A(t)/t,\mu-\epsilon). \label{eq:lambde} \eeq
where $\epsilon$ is a small, positive value, taken to be $\epsilon=0.0001$ in our simulations.\footnote{It is of course reasonable to assume that a receiver's delivery rate $\lambda$ will be less than the channel capacity $\mu$, but the reason for this explicit upper bound will become apparent in \eqref{eq:M}.} This is a practical way to choose $\lambda$ for our dynamic delivery model, because it allows the sender to dynamically adjust its transmission rate to accurately reflect its throughput and delay performance priorities. The stability of this system under inaccurate values of $\lambde$ is discussed in Section \ref{sec:lambdeanalysis}.

The dynamic delivery model also assumes that zero state delivery is the only method by which packets can be delivered to the receivers. It should be noted that this is a somewhat pessimistic performance estimate, as it does not take into account the effects of leader state and coefficient-based delivery studied in Sections \ref{analysis} and \ref{coeffbased}. However, since the effects of leader state and coefficient-based delivery were found to be difficult to predict, the zero state delivery delay provides a tractable upper bound on the expected delivery delay.

In many situations, zero state delivery provides a reasonable estimate of the delivery delay. For moderate to large values of $R$, leader state delivery has a relatively minor impact, and under coding schemes A and RLNC, the effects of coefficient-based delivery on the delay performance were relatively insignificant. Under
imperfect feedback conditions, the contributions from leader and coefficient-based delivery are further diminished. However, the zero state delivery model is not always accurate. If coding scheme B is used, the addition rate $\lambda \rightarrow \mu$, or $R$ the number of receivers is small, then leader state and coefficient-based delivery can still have a significant impact \grII{on the actual delivery performance.\footnote{Although in this section we \med{shall} use the zero state delivery model to make predictions about the receivers' throughput and delay performance, the performance measurements of Section \ref{sec:performance} are obtained from the combination of all three delivery methods.}}

\subsection{Decision metric}

Now that the dynamic delivery model has been established, we can outline how add and wait decisions are determined in the dynamic rate control scheme. Previously, we studied the baseline transmission schemes, where packets were added to the transmission queue with probability $\lambda<\mu$. By
contrast, in the dynamic rate control scheme we are about to introduce, a \emph{decision metric} $M$ is used to determine whether to add or wait. This allows the sender to dynamically adjust its addition rate based on the delivery performance of the receivers, resulting in better
throughput-delay performance. The sender calculates $M$ by weighing throughput peformance measures $P_T^A(r)$, $P_T^W(r)$ as well as delivery performance measures $P_D^A(r)$ and $P_D^W(r)$ for each receiver $r$, under the add and wait decisions respectively.

\subsubsection{Throughput performance}
Our first task is to measure the throughput performance of a receiver $r$ under the add and wait decisions.
Every time the sender waits, there is no increase in the total throughput. Therefore,
 \beq P_T^W(r)=0. \eeq
On the other hand, adding increases the total throughput by one packet, giving
 \beq P_T^A(r)=1. \eeq

\subsubsection{Expected time to zero state delivery}
To measure the delivery delay performance, we compare the expected delivery delay for a receiver in Markov state $k$, under add and wait decisions.
Our first task is to calculate the average time it takes for a receiver in Markov state $k$ to zero state deliver. This is equivalent to finding the average time to move from state $k$ to $0$ for the first time, under the Markov chain in Fig. \ref{markov}.

Let $E_k$ be the expected \emph{time to zero state}, i.e. expected number of time steps it takes for a receiver starting at state $k$ to reach 0 for the first time. Since our Markov chain is positive recurrent, we know that $E_k$ exists.

The receiver's journey from state $k$ to 0 can be considered as a series of traversals through the Markov chain from $k$ to $k-1$ for the first time, $k-1$ to $k-2$ for the first time, and so on. Because the transition probabilities between adjacent states in the Markov chain are the same for all $k>0$, the average time required for each traversal is the same. Therefore, the expected time to zero state starting from $k$ can be expressed as
 \beq E_k=k E_1. \label{eq:Ek}\eeq
 
Studying the Markov chain of Fig. \ref{markov}, there are three possible transitions at each time step. Starting at state $k$, the receiver's Markov state may increment, decrement or remain the same, with the probabilities listed in Table \ref{tab:markovtable} adapted to $\lambde$, as calculated in \eqref{eq:lambde}. Each state transition corresponds to a \grII{one-unit} increase in delay. From this information, we can establish the relationship between the time to the zero state for different values of $k$:
 \beq E_k=1+qE_{k-1}+pE_{k+1}+(1-p-q)E_k. \eeq
Substituting $k=1$, we obtain
 \beq E_1=1+qE_0+pE_2+(1-p-q)E_1. \eeq
As a starting condition, $E_0=0$, since in this case the receiver is already at $k=0$. Additionally from \eqref{eq:Ek}, $E_2=2E_1$. Substituting in these values, we obtain
 \beq E_1=\frac{1}{q-p}.\eeq
Therefore, using \eqref{eq:Ek} and the values in Table \ref{tab:markovtable} we obtain the result
 \beq E_k=\frac{k}{\mu-\lambda}. \eeq

\subsubsection{Delivery performance}
We are now in a position to determine $P_D^W(r)$ and $P_D^A(r)$, the expected time to zero state under add and wait decisions. For a receiver $r$ starting in state $k_r$, adding will result in one of two possible Markov states, $k_r +1$ and $k_r$, depending on whether the receiver experiences an erasure or not. Therefore, the expected time to zero state under adding is given by
 \beqa P_D^A(r)&=&\grII{\mubar E_{k_r+1}+\mu E_{k_r}}\nonumber \\
 &=& \frac{k_r+\mubar}{\mu-\lambda}. \eeqa
Similarly, waiting will result in receiver $r$ moving to state $k_r$ or $k_r -1$. Therefore the expected time to zero state under waiting is given by
 \beqa P_D^W(r)&=& \grII{\mubar E_{k_r}+\mu E_{k_r-1}}\nonumber\\
 &=& \frac{k_r-\mu}{\mu-\lambda}. \eeqa

\subsubsection{Benefits of adding and waiting}
For a given receiver $r$, we now calculate the benefit $B_W(r)$ and $B_A(r)$ of waiting and adding respectively.

\grII{Lower times to zero state are desirable, therefore we multiply the delivery performance of each receiver by a factor of -1. We also observe that when a receiver moves to the zero state, they will deliver \emph{all} the packets in their buffer. Therefore, we also scale the receiver's delivery performance by the number of undelivered packets $u(r)$ currently stored in their buffer. The information provided by $u(r)$ is particularly important because it tells us how many packets' delivery delays will be affected, and therefore how great an impact the sender's decision will have on the delay performance.}

On the other hand, throughput and delay may not be of equal importance. Therefore we scale the throughput performance by a weighting factor $f$, which determines the relative importance of one unit of throughput, compared with one unit of delay. This single free parameter $f$ is important becsuse it allows us to study the throughput delay sensitivity of the system. This idea is used in other work such as \cite{zeng12}, where the parameter $p$ effectively dictates the balance between the throughput and delivery delay.

Combining our results, we obtain the benefits of adding and waiting,
 \beqa B_A(r)&=&fP_T^A(r)-u(r)P_D^A(r) \nonumber\\
 &=& f-u(r)\frac{k_r+\mubar}{\mu-\lambda} \eeqa
 \beqa B_W(r)&=&fP_T^W(r)-u(r)P_D^W(r) \nonumber\\
 &=& -u(r)\frac{k_r-\mu}{\mu-\lambda}. \eeqa
 
\subsubsection{Decision metric}
We now combine these results to determine whether the sender should add or wait. Of course, in practice we are not able to determine the addition rate $\lambda$ in advance, so we substitute $\lambda$ with the estimate $\lambde$ from \eqref{eq:lambde}. For a single receiver $r$, the difference in performance between adding and waiting is given by
 \beqa M(r)&=&B_A-B_W\\
 &=&f-\frac{u(r)}{\mu-\lambde}.
 \eeqa
If $M(r)>0$, it indicates that adding is more beneficial to receiver $r$'s performance than waiting. It is interesting to note that $M(r)$ does not depend on the actual Markov state $k_r$ of the receiver. This comes about because in \eqref{eq:Ek}, the expected time to zero $E_k$ is linearly dependent on $k$. Adding a new packet merely increments the Markov state by 1, compared with waiting. Therefore the difference in performance between adding and waiting does not depend on the receiver's Markov state $k$.
 
Our decision metric is therefore given by the sum of the receivers' performance differences,
 \beqa M&=&\sum_{r=1}^R M(r)\nonumber\\
 &=& Rf-\frac{\sum_{r=1}^R u(r)}{\mu-\lambde}. \label{eq:M}
 \eeqa
If $M>0$ then the sender decides to add a new packet to the transmission queue. Otherwise, it waits. It is important to note that the $\lambda<\mu$ constraint set by \eqref{eq:lambde} ensures that the weighting of $u(r)$ will always be negative.

\subsection{Feedback variables}

There are two feedback variables \red{that} affect the decision metric, $M$: the total number of undelivered packets stored at the receivers, $\sum_{r=1}^R u(r)$, and the delivery rate estimate $\lambde$. We briefly discuss the impact of each variable on the sender's decision.

\subsubsection{Number of undelivered packets $\sum_{r=1}^R u(r)$}
Let us assume for now that $\lambde$ is constant. When this is the case, $u(r)$ becomes the only variable required to determine $M$. From \eqref{eq:M}, we can observe that there is a threshold $T_U$ number of undelivered packets stored \med{among the receivers} above which, $M<0$ and the sender will wait, but below which the sender will add. The threshold value can be found by solving the equation
 \beq Rf-\frac{T_U}{\mu-\lambde}=0, \eeq
yielding the solution
 \beq T_U=Rf(\mu-\lambde). \label{eq:undelivT}\eeq
Therefore the sender's decision strategy can be equivalently phrased as follows:\\\\
{\it If the total number of undelivered packets $\sum_{r=1}^R u(r)$ stored at the receivers is greater than or equal to $T_U$, then wait. Otherwise, add a new packet to the transmission queue.}\\\\
The greater the value of $T_U$, the longer the sender will spend adding new packets to the transmission queue, before waiting to reduce the number of undelivered packets at the receivers. This means that greater values of $T_U$ will generally result in higher throughput, but also higher delivery delays.

\subsubsection{Addition rate estimate, $\lambde$} \label{sec:lambdeanalysis}
Here we investigate the consequences of an inaccurate addition rate estimate, $\lambde$. Let us assume that there is some value $\lambda$, which is the correct addition rate for the system, and let $\lambde=\lambda+\Delta$ be the (possibly inaccurate) estimate based on the observed addition rate so far. The effect of the discrepancy $\Delta$ on the undelivered packet threshold of \eqref{eq:undelivT} is
 \beqa T_U'&=&Rf(\mu-\lambda-\Delta).\nonumber
 \eeqa
Notice that this only differs from \eqref{eq:undelivT} by a $\Delta$ term. The result is that the greater $\lambde$ is compared with $\lambda$, the smaller $T_U$ will be, and vice versa.

Interestingly, this results in a feedback loop where if $\lambde$ is too high, the addition rate will drop below $\lambda$, in turn reducing the measurement $\lambde$. On the other hand, a low value of $\lambde$ will have the opposite effect, raising the addition rate, and therefore the estimate $\lambde$. The end result is a stable addition rate, where fluctuations will to some degree be corrected by the system.
It can be observed in Fig. \ref{fig:lambde} that, in line with our analysis, $\lambde$ quickly converges upon a stable value.

\begin{figure}\figtopsmall\begin{center}
\includegraphics[width=\imgsize]{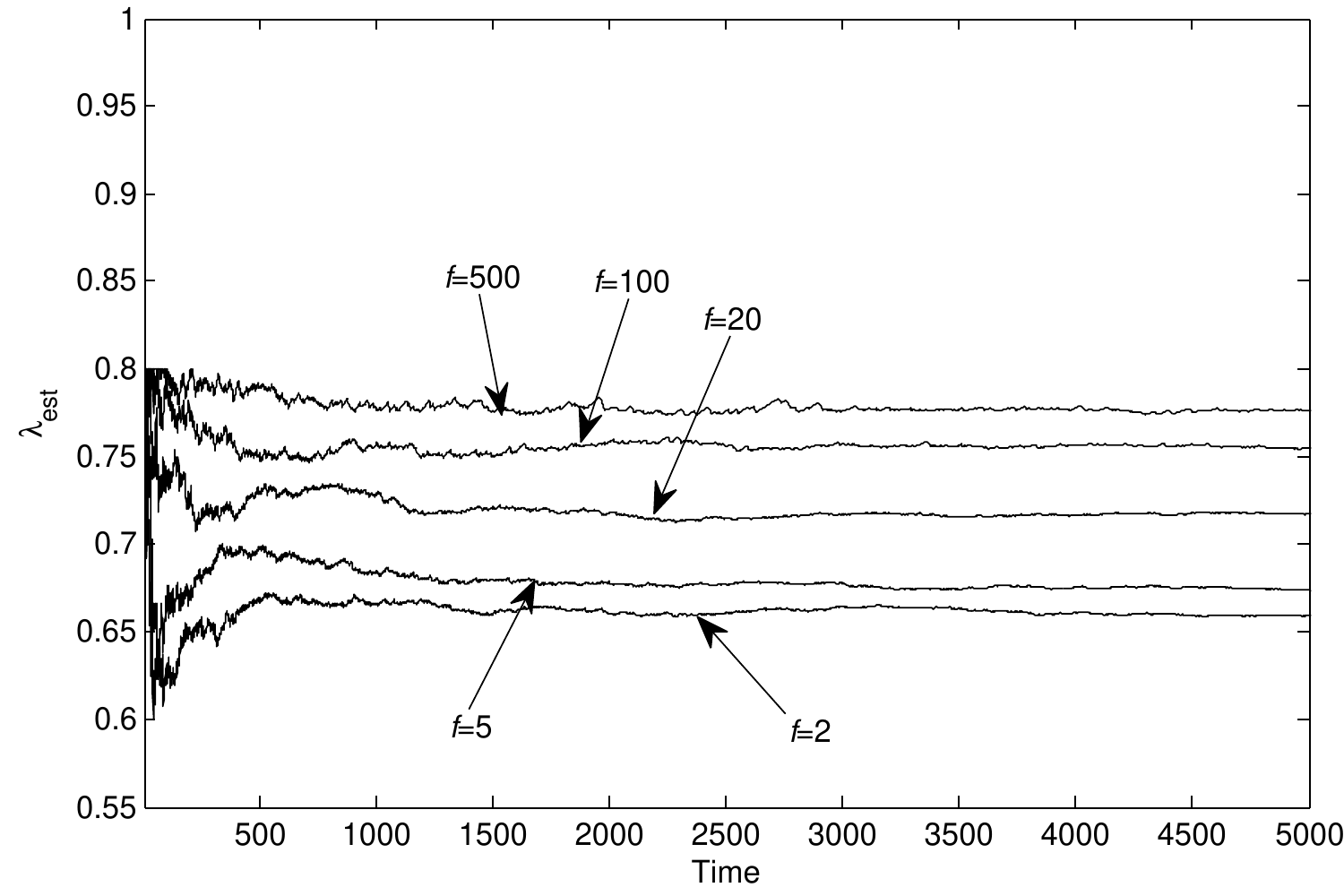}
\caption{The sender's estimated addition rate $\lambde$ as a function of time, for the weighting factors $f$ listed in the figure. The simulation was implemented with $R=4$ and $\mu=0.8$.} \label{fig:lambde}
\end{center}\figbottom \end{figure}
}
\section{Performance Comparison}\label{sec:performance}

\begin{figure}\figtopsmall\begin{center}
\includegraphics[width=\imgsize]{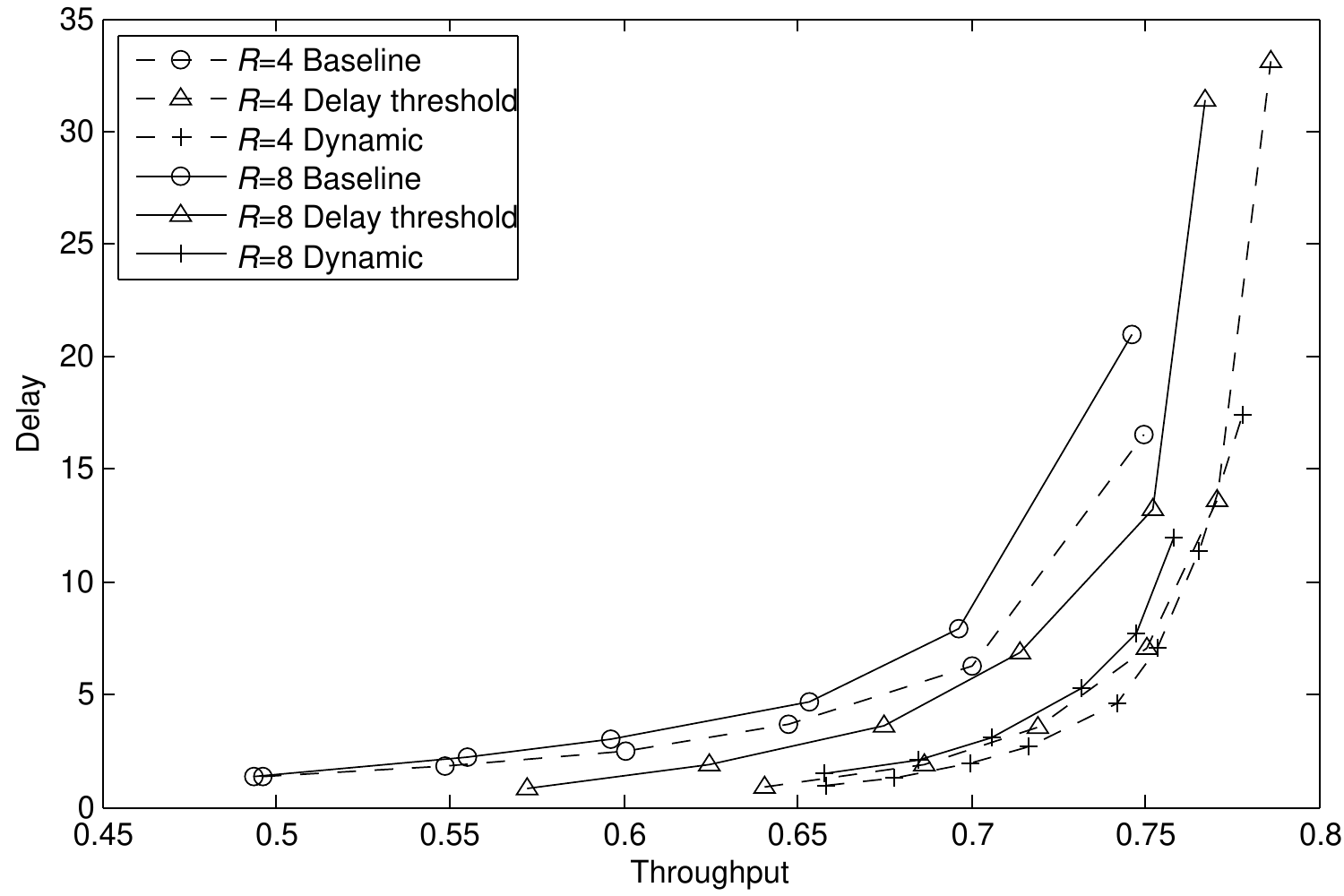}
\caption{\blue{Throughput-delay \grn{performance} for the coding scheme B transmission schemes. \bII{From left to right, the parameters for each transmission scheme are:}
Baseline: $\lambda=0.5,...,0.75$. Delay threshold: $T_D=2,...,100$. \bII{Dynamic: $f=2,...,500$.} For all transmission schemes $R=4$ and $\mu=0.8$.}}
\label{ratectrl}
 \end{center}\figbottom\end{figure}

Here we compare the performance of the \grn{three} \blue{coding scheme B transmission schemes.} Coding scheme B \grn{was chosen} since of the three \blue{coding} schemes studied in Section \ref{coeffbased} it has the best delay performance. In Fig. \ref{ratectrl} we compare the throughput-delay
performance of the \blue{coding scheme B transmission} schemes \blue{for 4 and 8 receivers}. The results are discussed here.

\subsection{Baseline rate control scheme}

As expected, \blue{the baseline rate control scheme exhibits the worst throughput-delay performance of the three rate control schemes. As the throughput increases, so does the average delivery delay, with the 4-receiver case performing marginally better than the 8-receiver case.}

\subsection{Delay threshold rate control scheme} \label{sec:delaythresh}
The delay threshold rate control scheme performs significantly better than the baseline rate control scheme. \grn{\med{The} strategy of} reducing the rate when one or more receivers is experiencing significant delays greatly improves the delivery delay. However, the delay performance for the 8-receiver case is significantly worse than the 4-receiver case. This is most likely caused by the transmission inefficiency of the stop mode.

In stop mode, the sender transmits an uncoded packet to allow the \grn{worst performing} receiver to deliver \med{its} next needed packet\bII{(s), thus improving \med{its} delivery delay}. However, in doing so \bII{every other receiver will incur a \grII{one-unit} throughput penalty}, since the uncoded transmission will not provide them with any innovative information. With larger numbers of receivers, this \bII{penalty can become quite significant.}

\subsection{Dynamic rate control scheme}\label{sec:dynamicperformance}
The dynamic rate control scheme further improves upon the delay threshold scheme. \bII{We can observe in Fig. \ref{ratectrl} that, as intended, increasing the weighting factor $f$ results in higher throughput. The real improvement over the delay threshold rate control scheme can be seen in the 8-receiver case, where for the same throughput, the dynamic rate control scheme experiences approximately half the delivery delay of the delay threshold scheme. This can be attributed to the fact it is a fairer, more well informed rate control scheme.

Unlike the delay threshold scheme, the dynamic rate control scheme does not disproportionately weight the needs of the worst performing receivers. When determining whether to add or wait, the sender weighs the requirements of \emph{all} receivers, instead of only the receiver(s) with the worst delay performance.

Furthermore, the dynamic rate control scheme considers both throughput and delay performance when determining whether to add or wait. This is in stark contrast to the baseline scheme, which only attempts to control the throughput, and the delay threshold scheme, for which the addition rate is controlled purely on the basis of \grII{the delay performance of the worst receiver}. By recognising that the add/wait decision is a tradeoff between throughput and delay performance, our dynamic rate control scheme is able to determine at each time slot which performance measure can most be improved upon under the current circumstances.}

\section{Conclusion}

We have demonstrated that the transmission rate of a broadcast
\blue{transmission} scheme can be dynamically \blue{adapted} to improve both
throughput and \blue{delivery} delay performance.

Analysing \blue{the baseline transmission schemes}, we used receivers' Markov states to
distinguish \med{among} three methods for \blue{packet delivery}: zero state, leader
state and coefficient-based delivery. We were able to accurately
model the zero state \blue{delivery} delay, and found that, in \blue{many} cases,
zero state delivery alone provided a reasonable approximation for
the expected delivery delay. Where there were more than a few
receivers, leader state delivery was observed to have a negligible
impact on the delivery delay. Although \blue{the RLNC scheme and coding scheme A} had only a small \blue{impact} on the \blue{delivery} delay, \blue{coding scheme B} resulted in significant improvements over zero
state delivery alone, by capitalising on more coefficient-based delivery opportunities.

Based on these observations we developed a dynamic rate adaptation scheme
that \grII{determined whether the sender should add or wait by comparing the benefit of each decision to the throughput and delay performance}. \grII{We found that this decision-making process was equivalent to regulating the sender's addition rate based on the total number of undelivered packets stored at the receivers. The}  dynamic rate adaptation scheme allowed noticeably better throughput-delay tradeoffs to be achieved, compared with existing approaches in the literature.

So far our work has only been in the context of \grII{receivers with homogeneous
channel rates}. While our analysis is equally applicable to heterogeneous
networks, a number of other issues including resource allocation and
fairness must also be considered. 


\section*{Acknowledgement}
The authors wish to acknowledge valuable comments and suggestions by the anonymous reviewers, which \grn{greatly} improved the presentation of this paper.

This work was supported under Australian Research Council Discovery Projects funding scheme (project no. DP120100160). The work of M. Medard was also supported by the Air Force Office of Scientific Research (AFOSR) under award No. FA9550-13-1-0023.
\bibliographystyle{IEEEtran}
\bibliography{IEEEabrv,../papers/bibtex}

\end{document}